\documentclass[a4paper,11pt]{article}
\usepackage{jcappub} 
\usepackage{amsfonts}
\usepackage{amsmath}
\usepackage{amssymb}
\usepackage{bm}
\usepackage{float}
\usepackage{array}
\usepackage{makecell}

\title{Detecting ionized bubbles around luminous sources during the reionization era using HI 21-cm signal}

\author[a]{Arnab Mishra,}
\author[b]{Chandra Shekhar Murmu,}
\author[a,1]{Kanan K. Datta%
\note{Corresponding author},}
\author[c]{Samir Choudhuri,}
\author[b]{Suman Majumdar,}
\author[a]{Iffat Nasreen,}
\author[a]{Sk. Saiyad Ali}
\affiliation[a]{Relativity \& Cosmology Research Centre, Department of Physics, Jadavpur University,
Kolkata 700032, India.}
\affiliation[b]{Department of Astronomy, Astrophysics \& Space Engineering, Indian Institute of Technology Indore, Indore 453552, India.}
\affiliation[c]{Centre for Strings, Gravitation and Cosmology, Department of Physics, Indian Institute of Technology Madras, Chennai 600036, India.}

\emailAdd{kanankdatta.physics@jadavpuruniversity.in}

\abstract{Measuring the properties of the intergalactic medium (IGM) and sources during the Epoch of Reionization (EoR) is of immense importance. We explore the prospects of probing the IGM and sources through redshifted 21-cm observations of individual ionized bubbles surrounding known luminous sources during the EoR.  
Accordingly, we simulate H{\footnotesize I} 21-cm maps, foreground contaminants, and system noise which are specific to the uGMRT and SKA1-Low observations. Following the subtraction of the foreground from the total visibility, we employ a visibility-based matched filter technique to optimally combine the desired H{\footnotesize I} 21-cm signal while minimizing the system noise.
Our analysis suggests that these ionized bubbles can be detected with more than $5 \sigma$ significance using approximately $\sim 2000$ and $\sim 3000$ hours of observation time with the uGMRT at redshift $7.1$ and $8.3$, respectively, when the mean neutral hydrogen fraction outside the targeted bubble is ${\sim 0.9}$. The SKA1-Low should be able to detect these bubbles with more than $8 \sigma$ significance using only $\sim 100$ hrs of observations. The total observing time increases both for the uGMRT and SKA1-Low when the mean neutral hydrogen fraction outside the targeted bubble decreases. Further, we investigate the impact of foreground subtraction on the detectability and find the signal-to-noise ratio decreases when smaller bandwidth is used. More importantly, we show that the matched filtering method can measure ionized bubble radius and constrain H{\footnotesize I}-neutral fraction reasonably well, providing deeper insights into the source properties and the intergalactic medium.}

\begin{document}
\maketitle
\flushbottom

\section{Introduction}
\label{sec:intro}
It is widely accepted that ionizing UV radiation from the first galaxies and quasars began to ionize the neutral hydrogen (HI) in the intergalactic medium (IGM) during the redshift range from $z \sim 20$ to $6$ \citep{Barkana_2001, Loeb+2010+i+vi, Pritchard_2012, bera2023}. During this process, fully ionized regions, commonly referred to as ionized bubbles, began to form around these early galaxies and quasars, marking the onset of the epoch of reionization (EoR). As time progressed, these ionized bubbles grew and eventually merged, resulting in the ionization of the entire universe~\citep{mellema2006, Furlanetto_2006, tirth2009}.\par

In recent times, there have been detections of several bright QSOs during the EoR~\citep{Mortlock_2011, Wu_2015, Ba_ados_2017, Wang_2018, Wang_2019, Matsuoka_2019a, Yang_2020, Wang_2021}, and a handful of bright galaxies have also been detected at high redshifts as well~\citep{Matsuoka_2019, witstok24}. These bright, luminous sources will produce large ionized bubbles (H{\footnotesize II} regions) around them within the IGM. At the initial and intermediate stages of reionization, isolated or nearly isolated ionized bubbles are expected, which will remain buried inside the neutral (or partially ionized) IGM. Observations of redshifted H{\footnotesize I} 21-cm signal are a promising tool to study the epoch of reionization, including the individual ionized regions \citep{ mellema2013, Majumdar_2011,majumdar2012, ghara2017, zack2020, giri2018, giri2018a,bianco2024,bianco2024a}. However, there are enormous challenges associated with these detections, as the signal is weaker by several orders of magnitude compared to the foreground contaminants~\citep{Di_Matteo+2002, Oh_2003, Santos+2005, Ali_2008}. Moreover, the system noise is much stronger than the desired H{\footnotesize I} signal.\par

Previous studies have explored the detection of H{\footnotesize II} regions in H{\footnotesize I} 21-cm images \citep{geil2008, mellema2013, Kakiichi_2017, giri2018, giri2018a,bianco2024,bianco2024a}. This method enables direct estimation of the H{\footnotesize I} neutral fraction, the sizes of ionized bubbles, and the properties of the underlying sources. However, direct imaging of H{\footnotesize I} using redshifted 21-cm signal during the Epoch of Reionization (EoR) is challenging with current instruments like the uGMRT, LOFAR, MWA due to their limited sensitivities. The SKA1-Low is anticipated to produce images of individual ionized bubbles using the H{\footnotesize I} 21-cm signal, but this requires considerably long observation times. 

This work investigates the prospects of individually detecting isolated ionized bubbles within the IGM, around known luminous sources using the \textit{matched filter} algorithm. This technique works when the system noise of the radio interferometer dominates over the targeted HI signal. But the functional form of the targeted signal is known. The idea of applying a matched filter technique was first proposed in \cite{Datta_2007}, which presents a visibility-based analytical framework to study the feasibility of detecting individual ionized bubbles during the EoR through radio-interferometric observations of redshifted H{\footnotesize I} 21-cm radiation. Later, it was shown that fluctuating intergalactic medium outside the targeted ionized bubble behaves as noise and hinders the detection of ionized bubbles of small sizes with radius $\lesssim 6$ Mpc \citep{Datta_2008}. Subsequent work such as \cite{Datta_2012} used detailed numerical simulations to study the detection prospects and associated challenges. \cite{Datta_2009} presents scaling relations, which enable us to quickly estimate detection prospects for various ionized bubble sizes, redshifts, and instruments. Further, \cite{Ghara_2020} used a Bayesian framework to explore the possibility of constraining the parameters that characterize the bubbles. This approach of detecting individual ionized bubbles in H{\footnotesize I} 21-cm maps is complementary to the widely followed approach of detecting the summary statistics of the signal such as the power spectrum, bispectrum, etc. While the statistical methods probe the large scale behavior of the reionizing sources and the intergalactic medium \citep{mesinger2016, majumdar2018, majumdar2020, trott2019, ghara2020, kamran2021, noble2024}, our approach sheds light on individual sources, ionized regions, and surrounding neutral medium. Interpreting the detection is also straightforward in our approach, unlike the statistical methods which might be complicated and model-dependent.\par

The majority of previous investigations have primarily focused on modeling H{\footnotesize I} 21-cm signal around ionized bubbles. In this work, we focus on targeted detection i.e., the location of the central luminous source is known from other observations. Along with simulated H{\footnotesize I} 21-cm maps which have a known source, we also simulate foreground contaminants and system noise. This is something that was not explicitly included in the previous works. Furthermore, we generated mock visibilities at baselines typical of the uGMRT/SKA1-Low observations. After subtracting the foreground from the total mock signal, we employ the matched filter technique on the residual signal. In that sense, the procedure adopted in this work resembles more with real observations. Moreover, we examine the constraints that observations with the uGMRT and SKA1-Low can impose on measuring bubble sizes and ionization fractions. \par

The outline of the paper is as follows: Section~\ref{sec:21-cm signal} provides a brief overview of the basics of the H{\footnotesize I} 21-cm signal around ionized bubbles. In section~\ref{sec:simulation}, we explain our methods to simulate the H{\footnotesize I} 21-cm signal around ionized bubbles, foreground contaminants, the generation of visibilities at baselines specific to a particular experiment, and system noise. Section~\ref{sec:ionized_bubble_detection} discusses foreground subtraction from the total visibilities and the application of the matched filter technique. Our results are presented in Section~\ref{sec:results}, while Section~\ref{sec:summary} presents a summary of the work undertaken in the paper alongside a discussion. Throughout our analysis we use the cosmological parameters $ h= 0.7$, $\Omega_{\rm m} = 0.27$, $\Omega_{\Lambda} = 0.73$,
$\Omega_{\rm b} = 0.044$, ${\sigma_8=0.83}$~ and ${n_{\rm s}=0.96}$ consistent with the WMAP measurements~\citep{Bennett_2013}.

\section{The H{\footnotesize I} 21-cm signal}
\label{sec:21-cm signal}
During the EoR, the ionizing radiation from the first generation of luminous sources, such as galaxies and quasars, leads to the emergence of H{\footnotesize II} regions (ionized bubbles) around these sources, and the remaining neutral (H{\footnotesize I}) medium surrounds these ionized regions. The surrounding H{\footnotesize I} medium is expected to emit a line emission with a rest-frame wavelength of approximately $21$ cm, which arises due to the hyperfine spin-flip transition in the neutral hydrogen atoms. This redshifted H{\footnotesize I} 21-cm signal emitted from these surrounding H{\footnotesize I} mediums can be observed as a differential brightness temperature against the Cosmic Microwave Background, and one can express this differential brightness temperature as \citep{Bharadwaj_2004},
\begin{equation}
{\rm \delta}T_{\rm b}({\bm n},z)=\frac{T_{\rm s} -T_\gamma}{1+z}\left(1-e^{-\tau_{\rm 21cm}}\right).
\end{equation}
$T_{\rm S}$, $T_\gamma$, and $\tau_{\rm 21cm}$ are the H{\footnotesize I} spin temperature, CMB temperature, and the optical depth for the H{\footnotesize I} 21-cm line emission respectively. All these quantities depend on redshift $z$ and the line of sight direction $\bm{n}$. Under the assumption that the H{\footnotesize I} 21-cm optical depth is small enough ($\tau_{\rm 21cm} \ll 1$) during the EoR, the above equation can be written as \citep{Bharadwaj_2005},
\begin{equation}
    {\rm \delta}T_{\rm b}(\bm{n}, z) = 4{\rm mK}\bigg(\frac{\rho_{\rm HI}}{\overline{\rho}_{\rm H}}\bigg)\sqrt{\frac{1+z}{\Omega_{\rm m}}}\bigg(\frac{\Omega_{\rm b}h^2}{0.02}\bigg)\bigg(\frac{0.7}{h}\bigg)\bigg(1 - \frac{T_\gamma}{T_{\rm s}}\bigg)\bigg[1 - \frac{1+z}{H(z)}\frac{\partial v}{\partial r}\bigg].
\end{equation}
$\rho_{\rm HI}$ and $\overline{\rho}_{\rm H}$ are the mass density of neutral hydrogen and the average mass density of total hydrogen (neutral and ionized). $\Omega_{\rm m}$ and $\Omega_{\rm b}$ are the matter and baryon density parameter, $h$ is the dimensionless Hubble parameter and $H(z)$ is the Hubble parameter at redshift $z$ respectively. The term inside the square brackets arises due to the redshift space distortion, and $\partial v/\partial r$ is the divergence of the peculiar velocity along the line of sight. Since the H{\footnotesize I} spin temperature of the neutral hydrogen in the IGM is expected to be coupled to the kinetic temperature of the gas during the EoR via the Lyman-$\alpha$ coupling~\citep{Wouthuysen+1952, Field+1958, Madau_1997}, it is expected to be much higher than the temperature of CMB ($T_{\rm s} \gg T_\gamma$). Therefore, one can usually neglect the term (1 - $T_\gamma/T_{\rm s}$) in the H{\footnotesize I} 21-cm differential brightness temperature expression during the EoR. It is easily seen that the differential brightness temperature of the H{\footnotesize I} 21-cm signal $\delta T_{\rm b}$ directly probes the neutral hydrogen content ($\rho_{\rm HI}$) in the IGM. In the absence of neutral hydrogen in the H{\footnotesize II} regions, this differential brightness temperature is expected to be zero. Therefore, this allows us to probe the individual ionized bubbles within the neutral IGM.\par
In this paper, we specifically focus on individual H{\footnotesize II} regions around high redshift bright quasars and galaxies like the ones reported in \cite{Mortlock_2011} and \cite{witstok24}, located at redshift $z=7.1$ and $z=8.3$ respectively. We expect large H{\footnotesize II} regions extending up to several tens of comoving Mpc around those bright luminous sources~\citep{Mortlock_2011, witstok24}. Our ultimate objective is to investigate the detectability of such individual ionized regions using radio interferometric observations of H{\footnotesize I} 21-cm signal. The following section presents mock simulations of similar ionized bubbles around very bright quasars and galaxies during the EoR.

\section{Simulating mock data}
\label{sec:simulation}
To investigate the prospects of detecting ionized bubbles individually in H{\footnotesize I} 21-cm maps, we first simulate mock data that resembles the uGMRT and SKA1-Low observations and at observational frequencies relevant to the EoR. It consists of simulations of the H{\footnotesize II} region around bright quasars/galaxies in large-scale H{\footnotesize I} maps, foreground contaminants, and system noise from radio interferometers. Subsequently, these simulated maps are converted into visibilities that mimic the uGMRT/SKA1-Low-like observations.\par
Our simulations are centered around two different observing frequencies, $175$ MHz and $153$ MHz, corresponding to the redshifts of $z \approx 7.1$ and $8.3$, respectively, for the H{\footnotesize I} 21-cm line emission. The selection of these two frequencies is motivated by the discoveries of a bright QSO at redshift $z=7.1$~\citep{Mortlock_2011} and a large ionized bubble at redshift~$z=8.3$~\citep{witstok24}. The angular size of our simulation cube is set to $\sim 3.3^\circ \times 3.3^\circ$ which is closer to the uGMRT/SKA1-Low primary field of view at the observing frequencies considered here. Our simulation box has a total of $512$ grids on each side with a resolution of $1.03$ Mpc and a total comoving length of $527$ Mpc on each side. This results in an angular resolution of $\sim 23^{\prime\prime}$ and a total angular size of $3.3^\circ$. Here, we have used up to $256$ frequency channels with a total bandwidth of $16$ MHz, resulting in a channel width of $\Delta \nu= 62.5$ kHz.

\subsection{H{\footnotesize I} 21-cm signal around ionized bubbles}
Here, we briefly describe simulated ionized regions surrounding a very bright quasar at $z = 7.1$, and around a collection of galaxies at $z = 8.3$ during the EoR. First, we simulated the dark-matter distribution using the  $N$-body code\footnote{\url{https://github.com/rajeshmondal18/N-body}} developed by \citep{Bharadwaj_2004a}.
We simulated this within a cubic box of the comoving size of $\sim 527$ Mpc with $2048^3$ grids and the same number of particles. The minimum halo mass resolved in this simulation is ${\sim6.26\times 10^{9} M_\odot}$, corresponding to a minimum of $10$ dark matter particles, each with a mass of ${\sim6.26\times 10^{8} M_\odot}$. \textbf{We note that our simulations do not include the contributions of low-mass halos in the range $\bm{10^{8} M_\odot}$ to $\bm{\sim6.26 \times 10^{9} M_\odot}$. In reality, these halos are expected to play a significant role in the reionization of the IGM, potentially giving rise to more diffuse, complex, and smaller ionized structures during the EoR.} Then we use a Friends-of-Friends~\footnote{\url{https://github.com/rajeshmondal18/FoF-Halo-finder}} (FoF)~\citep{Mondal_2015} halo finder to identify the collapsed halos that host the galaxies/bright quasars. These galaxies/quasars serve as the sources of ionizing photons during the EoR. Subsequently, the resolution of simulated cubes is degraded to $512^3$ grids. This resolution resembles the corresponding angular resolution of the uGMRT and SKA1-Low and has the added benefit of reduced computational requirements for simulation. We then employ a semi-numerical prescription \citep{Choudhury_2009, Majumdar_2014, Mondal_2017} on these outputs to simulate ionization maps and the H{\footnotesize I} 21-cm differential brightness temperature maps.

The semi-numerical approach compares the average number of ionizing photons $\langle n_\gamma\rangle_R$ and neutral hydrogen atoms $\langle n_{\rm H}\rangle_R$ within a spherical region of radius $R$ around a grid cell. If the ionization condition $\langle n_\gamma \rangle_R \geq \langle n_{\rm H}\rangle_R$ is met for any radius $R$, then the grid cell is reckoned to be ionized, within that region. Otherwise, an ionized fraction value of $x_{\rm HII} = \langle n_\gamma \rangle_{\rm grid} / \langle n_{\rm H} \rangle_{\rm grid}$ is assigned to that grid cell. This procedure is repeated for varying radius $R$ up to a maximum value $R_{\rm max}$, over the whole simulation box. It is assumed that the number of ionizing photons from halos, $N_\gamma$ contributed to the ionization of the IGM, follows the relation $N_\gamma \propto M_{\rm h}$, where $M_{\rm h}$ is the mass of the corresponding halo.
\par
\begin{figure}[htbp]
\centering
\includegraphics[width=0.49\textwidth]{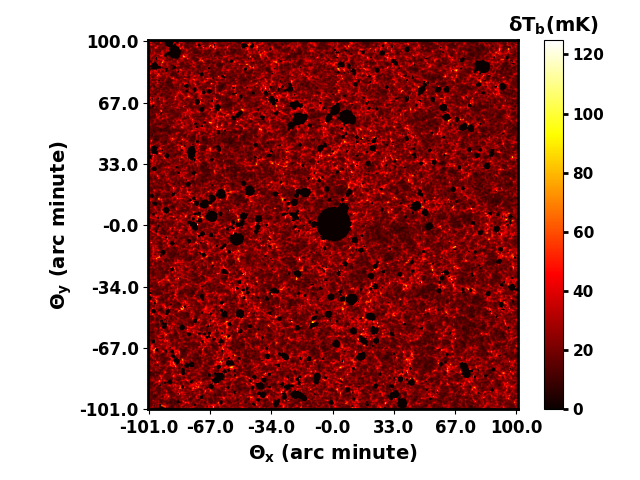}
\includegraphics[width=0.49\textwidth]{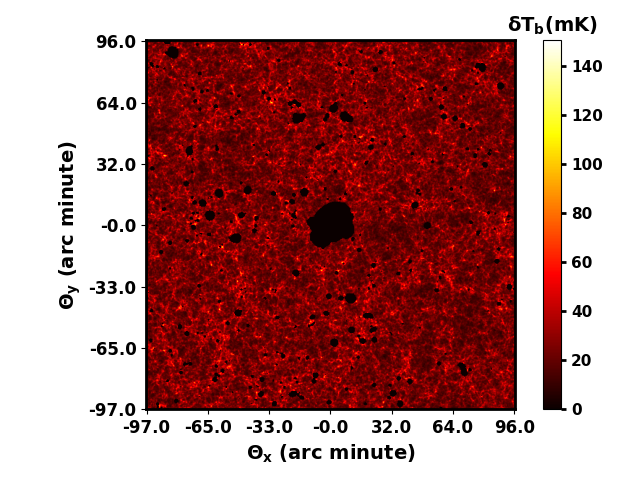}
\includegraphics[width=0.49\textwidth]{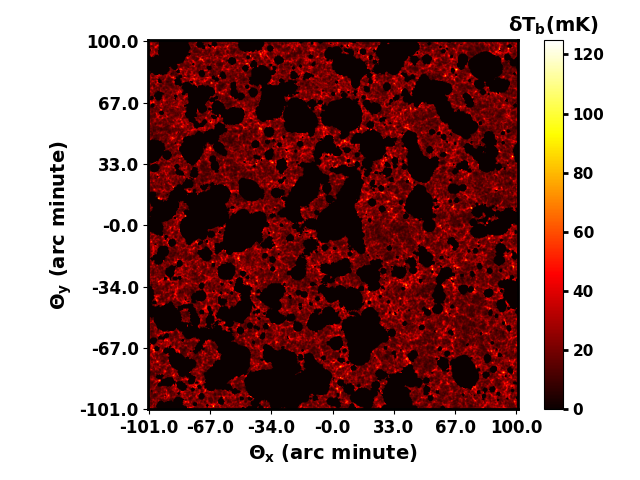}
\caption{The top left and bottom panels show differential brightness temperature maps of the H{\footnotesize I} 21-cm signal around an ionized bubble of comoving size $23.5$ Mpc at $z=7.1$ corresponding to observing frequency of $175$ MHz. The top right panel shows a similar map around an ionized bubble of comoving size $27.8$ Mpc at $z=8.3$ corresponding to observing frequency of $153$ MHz. The mass-averaged neutral hydrogen fraction resulting from the galaxies outside the central H{\footnotesize II} region is $0.88$, $0.94$, and, $0.52$ for the top-left, top-right, and bottom panels respectively. The angular size of simulated maps is $\sim 3.3^{\circ}\times 3.3^{\circ}$ with $ \sim 23^{\prime\prime}$ resolution.}
\label{fig:quasar_bubble}
\end{figure}

Here, we consider two scenarios. First, we aim to simulate an ionized region around a very bright quasar at a redshift of $7.1$ similar to the one reported in \cite{Mortlock_2011}. In the second case, we simulate an ionized region around a collection of galaxies at redshift $8.3$ as reported in \cite{witstok24}. In the first scenario, we assume that such a very bright quasar is hosted by the most massive halo (of mass $M_{\rm h} \sim 6\times10^{12}, M_\odot) $ found in our simulation \citep{majumdar2012}. We further assume that the ionizing photon emission rate by the QSO is $\sim 1.3\times10^{57} \, {\rm s}^{-1}$ \citep{Datta_2012, majumdar2012}. We note that along with the central quasar, many galaxies reside in and outside the QSO ionized regions. These galaxies are also contributing ionizing photons. The mass-averaged neutral Hydrogen fraction resulting from galaxies outside the QSO H{\footnotesize II} region is $\langle x_{\rm HI} \rangle \sim 0.88$ at a redshift of $z=7.1$. In figure~\ref{fig:quasar_bubble} (top left panel), the resulting H{\footnotesize I} 21cm map is shown with the ionized bubble around the quasar in the center of the map. The ionized region around the central bright QSO is considerably bigger (of radius around $23.5$ comoving Mpc) than all other ionized regions produced by the galaxies.\par
In the second case, we consider a collection of galaxies near the most massive halo (of mass $M_{\rm h} \sim 3.5 \times 10^{12} M_\odot$) in our simulation at $z=8.3$. Simulation specifications such as the cubic box size, total number of particles used, and number of grids along a direction are the same as above. The ionizing luminosity of these galaxies, including the most massive halo in our simulation, is adjusted to create an H{\footnotesize II} bubble with a radius of $\sim 27.8$ Mpc, and the resulting mass-averaged neutral fraction is around $\langle x_{\rm HI}\rangle \sim 0.94$. The ionized bubble from this scenario is shown in the top right panel of figure~\ref{fig:quasar_bubble}. We see that the ionized bubble in this scenario is slightly aspherical in shape, mainly because it is created by a collection of sources located at different locations inside the bubble.\par

In the above two cases, the targeted H{\footnotesize II} bubbles are isolated and nearly spherical. This geometry arises because the ionizing radiation from other sources is negligible, and the IGM outside the targeted bubbles remains highly neutral. While these scenarios may be suitable at the onset of reionization, they are likely unrealistic for the later stages of the process. At later stages of reionization, ionized bubbles around very bright sources are surrounded by neighboring ionized regions, leading to an aspherical shape for the targeted bubble. The third scenario accounts for this complexity. The bottom panel of Fig. \ref{fig:quasar_bubble} displays a 2D map of the differential brightness temperature (${\delta T_b}$) at redshift $z=7.1$, where the central targeted H{\footnotesize II} bubble appears highly aspherical. The quasar at the center is the same as the first case. Surrounding it there are numerous other H{\footnotesize II} regions, unlike the first case. This results in a much lower mean neutral fraction, which is $0.52$ in this scenario.\par

We want to mention the considerable uncertainty regarding the ionizing photon emissivity rate from QSO, and therefore, the photon emissivity rate and the resulting QSO H{\footnotesize II} region may vary. The consequences of this uncertainty will be discussed later in the paper.

\subsection{Foreground simulations}
The astrophysical foregrounds that contaminate the EoR H{\footnotesize I} 21-cm signal observations comprise various components \citep{Jeli__2008}, including diffuse galactic synchrotron emission, radio synchrotron emission from point-like sources, free-free radio emission from diffuse ionized gas, etc. 

\begin{figure}[htbp]
\centering
\includegraphics[width=0.49\textwidth]{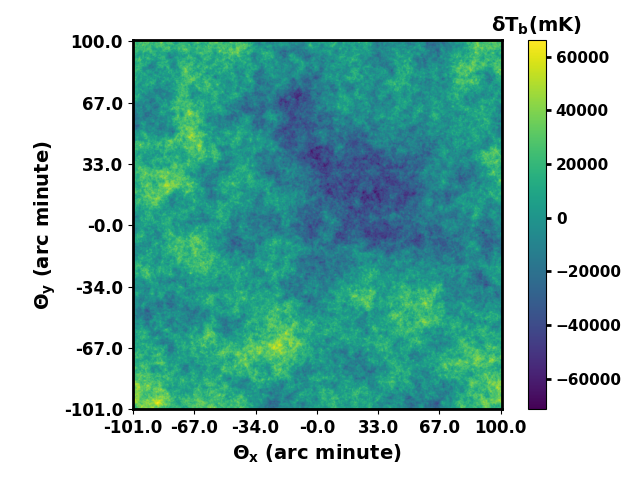}
\includegraphics[width=0.49\textwidth]{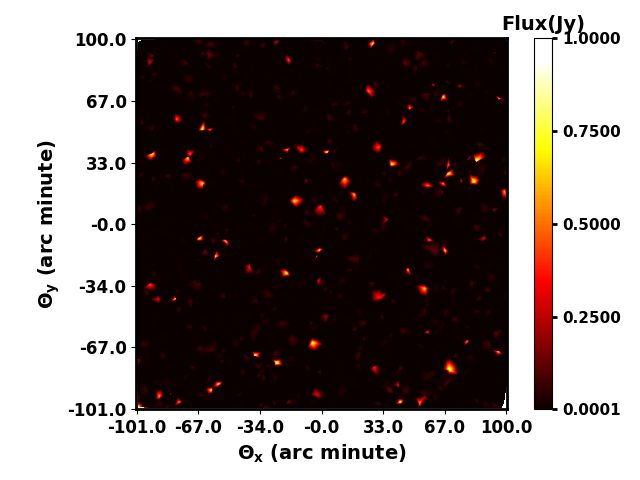}
\caption{The left panel shows a simulated 2D image of the diffuse galactic synchrotron emission (DGSE) of angular size $\sim 3.3^{\circ}\times 3.3^{\circ}$ with resolution of $\sim 23^{\prime\prime}$ at frequency $175$ MHz. The right panel shows a 2D image of extragalactic radio point sources with a similar angular size, resolution and observing frequency.}
\label{fig:dgse}
\end{figure}

The diffuse galactic synchrotron emission (DSGE) is one of the most dominant foreground components.
We model the angular power spectrum corresponding to the diffuse galactic synchrotron emission using a power-law formulation \citep{Choudhuri_2014} as,
\begin{equation}
{C_l}^M(\nu) = A_{150}\times {\left(\frac{1000}{l}\right)}^{\beta} {\left(\frac{\nu_0}{\nu}\right)^{2\alpha}}.
\end{equation}
Here, \(\nu_0 = 150\) MHz and \(\nu = \nu_0 + \Delta\nu\). The power-law index, \(\beta\), is set to $2.34$, and the amplitude \(A_{150} = 513 \, \text{mK}^2\) is based on previous work \cite{10.1111/j.1365-2966.2012.21889.x}. The parameter $\alpha$ describes how the power spectrum changes with frequency, which is set to $2.8$ \citep{Santos+2005, Ali_2008}. This simplified model allows us to simulate DGSE maps and to study their impact on the detection of individual ionized bubbles in the H{\footnotesize I} 21-cm maps. We use the following equation to generate Fourier components of the DGSE on each gridded baseline $U$, which is given by

\begin{equation}
\Delta\tilde{T}(U) = \sqrt{\frac{\Omega {C_l}^M}{2}}[x(U)+iy(U)].
\end{equation}
Here, $\Omega$ denotes the total solid angle corresponding to the spatial size of simulated maps. $x(U)$ and $y(U)$ are independent Gaussian random variables with zero mean and unit variance. Upon simulating 2D DGSE maps in Fourier space, we apply the inverse Fourier transform on them to obtain the simulated sky map $\delta T_b (\bm{\theta})$   at a specific frequency. Subsequently, we use the frequency scaling to obtain the DGSE map at other desired frequencies. Figure \ref{fig:dgse} shows a single realization of the 2D simulated DGSE map centered at frequency $\nu=175$ MHz corresponding to redshift $z=7.1$. As mentioned earlier, the angular resolution of these maps is $23^{\prime\prime}$, and the strength of temperature fluctuations changes with angular resolution. We note that the mean temperature of the DGSE map is set to zero, as the radio interferometric experiments, such as the uGMRT/SKA1-Low, are not sensitive to the mean due to the lack of the zero spacing baseline. Subsequently, we obtain mock sky maps at multiple frequency channels by combining the H{\footnotesize I} 21-cm and the foreground maps simulated above.

\subsubsection{Extragalactic radio point sources}
\label{subsubsec:ps}
One of the most significant foreground contaminants in H{\footnotesize I} 21-cm signal observation arises from the extragalactic radio point sources, such as Active Galactic Nuclei (AGN), quasars, and radio galaxies. At lower frequencies (${\sim 150}$ MHz), these sources dominate the sky at smaller angular scales $\lesssim 4^{\circ}$ \cite{Ali_2008}. Therefore, in order to study the H{\footnotesize II} regions at frequencies of $150$ and $175$ MHz, it is essential to model the point sources efficiently.

In our simulations, we model the population of extragalactic radio point sources using the following $150$ MHz differential source count model from~\cite{10.1111/j.1365-2966.2012.21889.x}:
\begin{equation}
    {\frac{dN}{dS}=\frac{10^{3.75}}{\rm Jy.Sr}\left(\frac{S}{1\,\rm Jy}\right)^{-1.6}},
    \label{eq:diff_sorce_count}
\end{equation}
where $S$ is the flux density of the point sources. We consider the point sources to be in the flux range between $0.1$ mJy to $1$ Jy
and they are randomly distributed across an angular size of ${\sim 3.3^{\circ}\times 3.3^{\circ}}$. Based on these parameters, we obtained $7775$ point sources from the simulation. We model the spectral energy distribution of the point sources using a simple power law (${S\propto \nu^{-\alpha}}$), where the spectral index $\alpha$ for each source is randomly assigned within the range of $0.7$ to $0.8$. The right panel of Figure \ref{fig:dgse} shows a 2D map of the point sources randomly distributed across an angular area of approximately $3.3^{\circ}\times 3.3^{\circ}$, with an angular resolution of ${\sim 23^{\prime\prime}}$.

\subsection{Visibilities for spherical ionized bubbles: An analytical approach}

In radio interferometric observations, the measured quantity is the visibility, denoted as $V(\bm{U},\nu)$, which can be written as the Fourier transform of the product of the specific intensity pattern of the sky, $\delta I_{\nu}(\bm{\theta})$ and $A(\bm{\theta}, \nu)$. It is given as,
\begin{equation}
    V(\bm{U},\nu)=\int d^2\theta \, A(\bm{\theta}, \nu)\,\delta I_{\nu}(\bm{\theta})\,e^{2\pi i\bm{\theta}.\bm{U}}.
    \label{eq:visb_sp}
\end{equation}
Here, the baseline $\bm{U}=\bm{d}/\lambda$ denotes the antenna separation $\bm{d}$ projected on the plane perpendicular to the line of sight in units of the observed wavelength ($\lambda$) of the signal. $\bm{\theta}$ is a two-dimensional vector on the plane of the sky with the origin at the center of the field of view (FoV), and $A(\bm{\theta}, \nu)$ is the beam pattern of the individual antenna. Assuming uniformly illuminated circular aperture of diameter $D$, we model the antenna primary beam pattern as \citep{Choudhuri_2014},
\begin{equation}
    A(\bm{\theta}, \nu)=\left[\left(\frac{2\lambda}{\pi \theta D}\right)J_1\left(\frac{\pi \theta D}{\lambda}\right)\right]^2,
\end{equation}
where $J_1$ is the first order Bessel function. The antenna diameter $D=45$m and $35$m for the uGMRT and SKA1-Low respectively. The specific intensity pattern of the sky, $\delta I_{\nu}(\bm{\theta}) \equiv \delta T_{\rm b}({\bm{n}},z)$ and can be written as,
\begin{equation}
\delta I_{\nu}({\bm{n}},z) = \frac{2k_B\nu^2}{c^2} {\rm \delta}T_{\rm b}({\bm{n}},z).
\label{eq:sp_int}
\end{equation}
Considering a spherical ionized bubble surrounded by a uniform H{\footnotesize I} background at the center of the field of view, we can derive the analytical formula for the visibility of the H{\footnotesize I} 21-cm signal at a particular frequency channel $\nu$ and baseline $\bm{U}$ following~\cite{Datta_2007} as,
\begin{equation}
      S_{\rm center}(\bm{U},\nu)=-\pi \bar{I_\nu} x_{\rm HI} {\theta_{\nu}}^2 \left[\frac{2J_1(2\pi U \theta_{\nu})}{2\pi U \theta_{\nu}}\right] \Theta\left(1 - \frac{\left|\nu - \nu_c\right|}{\Delta\nu_b}\right),
      \label{eq:vis_HI}
\end{equation}
where
\begin{equation}
\bar{I_\nu}=2.5\times10^2 {\rm Jy/sr}\left(\frac{\Omega_bh^2}{0.02}\right)\left(\frac{0.7}{h}\right)\left(\frac{H_0}{H(z)}\right)
\end{equation}
is the specific intensity from uniform H{\footnotesize I} background at redshift $z$, $x_{\rm HI}$ is the mass-averaged neutral hydrogen fraction, $\theta_R$ is the angular size of the ionized bubble corresponding to comoving radius $R$, $J_1$ is first order Bessel function and $\Delta\nu_b$ is ionized bubble size in frequency. Note that a spherical ionization bubble will appear as a circular one at a particular frequency channel. We use this analytical equation to design a filter for detecting the ionized bubbles in the mock H{\footnotesize I} 21-cm maps obtained from simulations. 

\subsection{Simulating visibilities}

\begin{figure}[htbp]
\centering
\includegraphics[width=0.49\textwidth]{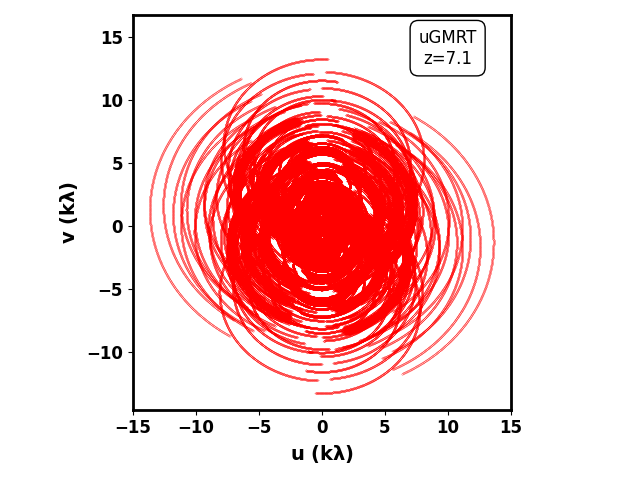}
\includegraphics[width=0.49\textwidth]{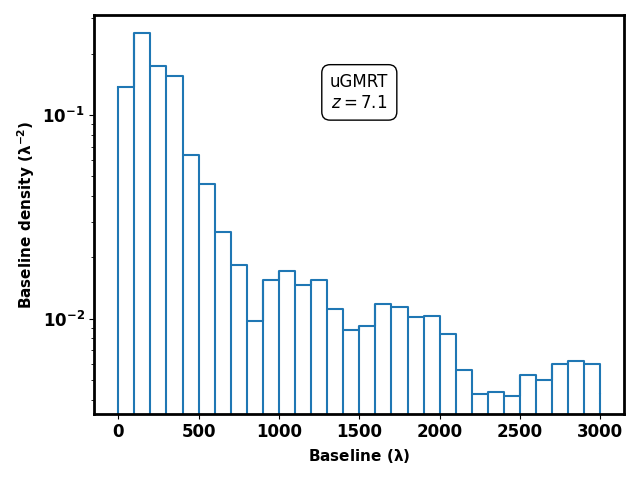}
\\

\includegraphics[width=0.49\textwidth]{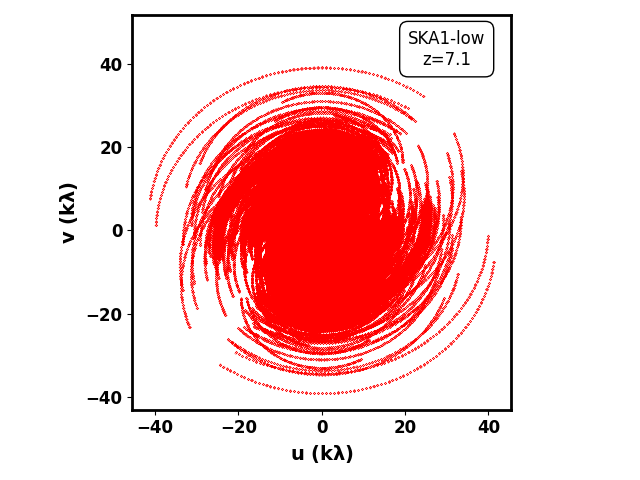}
\includegraphics[width=0.49\textwidth]{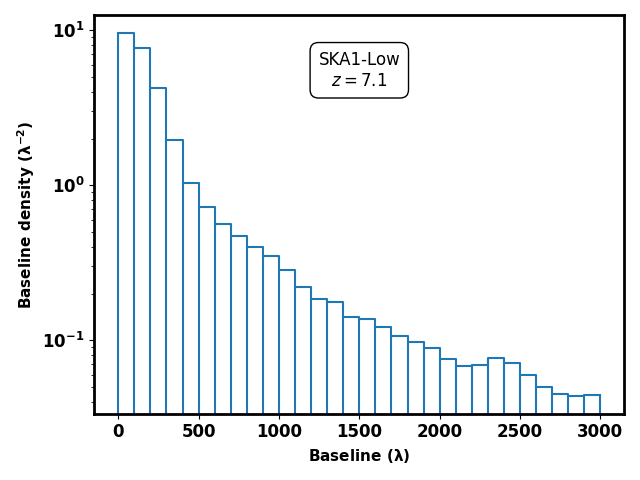}
\caption{Left panels show baseline coverage for $8$ hours of the uGMRT (upper left panel ) and SKA1-Low (lower left panel) observations at $175$ MHz. Right panels show corresponding baseline density as a function of baseline.}
\label{fig:baseline-dis}
\end{figure}

To simulate mock uGMRT and SKA1-Low observations, we consider baseline distributions corresponding to both instruments. There are currently $30$ active antennas at uGMRT \citep{2017CSci..113..707G} and the proposed number of antennas in SKA1-Low is $512$ \citep{ska_telescope}. Table \ref{tab:Baselines} summarises instrument specifications. Figure \ref{fig:baseline-dis} shows the baseline coverage for an $8$-hour uGMRT and SKA1-Low observations at the $175$ MHz frequency corresponding to $z=7.1$ for the H{\footnotesize I} 21-cm signal. It also shows the baseline density i.e., the total number of baselines per unit area of circular annulus corresponding to baseline $U$ and $U+dU$. For the uGMRT, baselines vary from $32\lambda$ to $13500\lambda$ and for the SKA1-Low, the baseline range extends from $4\lambda$ to $37000\lambda$ over the same $8$ hour, $175$ MHz observations. As expected, the number density of baselines and their extent are much greater for the SKA1-Low, as it has more antennae, and some of them are placed at large distances compared to the uGMRT. The baseline distribution at $153$ MHz (corresponding to $z=8.3$) is very similar to that at $175$ MHz but will be reduced (refer to Table \ref{tab:Baselines}).\par
\begin{table}[htbp]
    \centering
 
    \begin{tabular}{|c|c|c|c|c|c|c|}
    \hline
    Observation & \makecell{No. of \\ antennae} & $z$ & $\nu$(MHz) & \makecell{No. of baselines \\ per snapshots}  & $U_{max}(\lambda)$ & $U_{min}(\lambda)$   \\
    \hline
    uGMRT & $30$ & $7.1$ & $175$ & $435$ & $13500$ &$32$ \\ 
    \hline
    uGMRT & $30$ & $8.3$ & $153$ & $435$ & $12000$ &$28$ \\
    \hline
    SKA1-Low & $512$ & $7.1$ & $175$ & $130816$ & $37000$ &$4$ \\
    \hline
    SKA1-Low & $512$ & $8.3$ & $153$ & $130816$ & $32000$ &$3.5$\\ \hline
    \end{tabular}
    \caption{This shows some relevant parameters for the uGMRT and SKA1-Low at different observing frequencies.}
    \label{tab:Baselines}
\end{table}

We generate the mock visibility data of the H{\footnotesize I} 21-cm signal and the foregrounds (DGSE and extragalactic point sources) at all simulated baselines at $256$ frequency channels, centered around frequencies $\nu=175$ MHz and $153$ MHz, using the Discrete Fourier Transformation (DFT) of the total simulated maps. For simplicity, we keep the baseline distribution the same in all frequency channels.

\begin{figure}[htbp]
    \centering
    \includegraphics[width=0.46\textwidth]{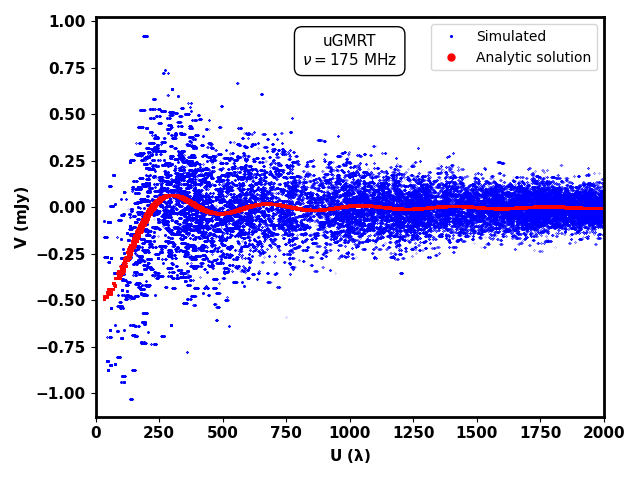}
    \includegraphics[width=0.49\textwidth]{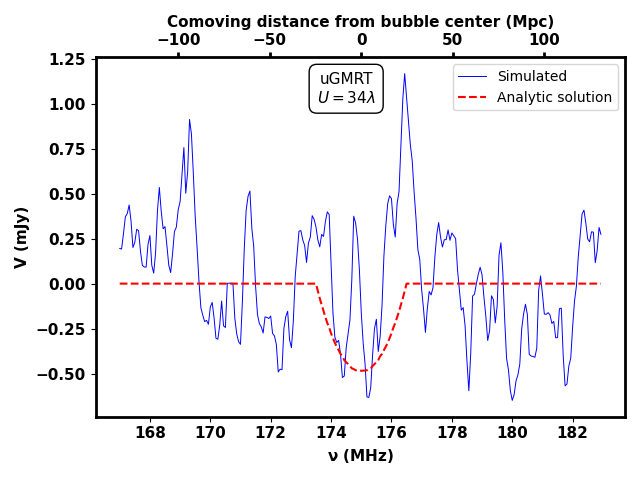}
    \\
    \includegraphics[width=0.46\textwidth]{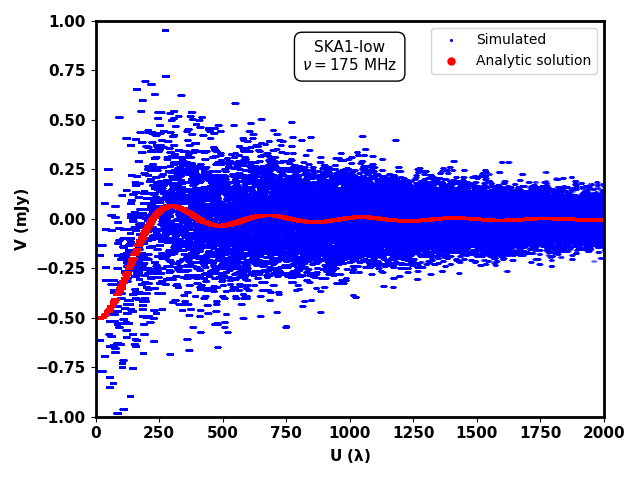}
    \includegraphics[width=0.49\textwidth]{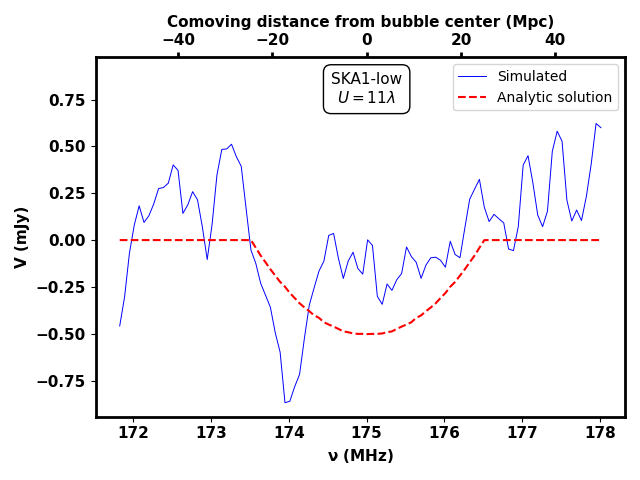}
    \caption{ Left panels show simulated visibilities as a function of baseline $U$ for the central frequency channel at $175$ MHz for the uGMRT (upper panel) and SKA1-Low (lower panel). Red lines show analytical predictions for a spherical ionized bubble of the same radius as in simulation and uniform H{\footnotesize I} outside the bubble. Right panels show the same as a function of observing frequency $\nu$ for fixed baselines $U=34 \lambda$ (uGMRT) and $U=11 \lambda$ (SKA1-Low).
    }
    \label{fig:vis1}
\end{figure}

Figure \ref{fig:vis1} shows simulated visibilities corresponding to the 2D central H{\footnotesize I} 21-cm map around the ionized bubble centered at $z=7.1$. The top left and top right panels show plots of the signal as a function of baseline and frequency respectively, for the uGMRT and the bottom panels display results for the SKA1-Low. We also compare the simulated visibility with the analytical estimation assuming a perfectly spherical ionized bubble of comoving radius $23$ Mpc and surrounded by uniform and fully neutral medium (refer to equation \ref{eq:vis_HI}). The IGM around the ionized bubble in simulated maps is not uniform, but traces the underlying DM distribution and is also filled with many small ionized bubbles. This causes H{\footnotesize I} density to fluctuate and consequently, the visibility signal also fluctuates as a function of baseline and frequency. We see in Figure \ref{fig:vis1} that the H{\footnotesize I} fluctuations are stronger at lower baselines which reflects the fact that H{\footnotesize I} power spectrum $P_{\rm HI}(k)$ (or equivalently the visibility-visibility correlations) increases at lower $k$-modes or baselines \citep{ali2005, Datta_2008}. We further see that the fluctuating H{\footnotesize I} outside the bubble dominates over the visibilities solely due to the H{\footnotesize I} 21-cm signal around a spherical ionized bubble which is buried in a uniform H{\footnotesize I} background. Visibilities are similar for the second scenario at $z = 8.3$ for both experiments and we do not show them here explicitly. 

\subsection{System noise}
To make our study more realistic, we introduce system noise contribution from radio interferometers in our simulated mock visibility data. The system noise contribution to the measured visibility in each baseline and frequency channel is an independent Gaussian random variable with zero mean and the root mean square \citep{Datta_2007}

\begin{equation}
\sqrt{\langle N^2 \rangle} = \frac{\sqrt{2}k_BT_{\rm sys}}{A_{\rm eff}\sqrt{\Delta\nu\Delta t}},
\label{eq:sys-noise}
\end{equation}
where $T_{\rm sys}$ represents the total system temperature, $A_{\rm eff}$ is the effective collecting area of the antenna, and $\Delta\nu$ and $\Delta t$ denote the channel width and correlator integration time, respectively.

For the uGMRT, $T_{\rm sys} = 300$ K and $A_{\rm eff}/(2k_{\rm B}) = 0.33$ K/Jy \citep{2017CSci..113..707G} at frequency $\nu \sim 175$ MHz. The estimated root mean square noise in visibility is approximately $0.64 \, {\rm Jy}$ for frequency channel width ($\Delta \nu$) of $62.5$ kHz and integration time ($\Delta t$) of $16$ seconds in each baseline. $T_{\rm sys} = 300$ K and $A_{\rm eff}/(2k_{\rm B}) = 0.348$ K/Jy for the SKA1-Low at the same frequency considering antenna efficiency factor unity. $T_{\rm sys} $ and $A_{\rm eff}/(2k_{\rm B})$ change to $400$ K and $0.348$ K/Jy for the SKA1-Low at $153$ MHz frequency. For the uGMRT, these numbers are $400$ K and $0.33$ K/Jy \citep{2017CSci..113..707G} respectively at $153$ MHz frequency. We note that the rms noise will be reduced by a factor of $\sqrt{t_{\rm obs}/\Delta t}$ where $t_{\rm obs}$ is the total observation time.\par
Our mock simulation considers $8$ hrs of observations with $16$s and $320$s integration time for the uGMRT and SKA1-Low respectively. This produces a total of $783000$ and $11773440$ baselines for the uGMRT and SKA1-Low respectively. Lower integration time will produce a higher number of baselines for the SKA1-Low and hence will increase the requirement of computational resources which is beyond our reach. The system noise contribution for $8$ hrs of observations will be too high to detect the H{\footnotesize I} 21-cm signal around the individual ionized bubbles.
Therefore, we present our results for $2048$ and $3072$ hrs of uGMRT observations at redshifts $7.1$ and $8.3$, respectively, assuming a total of $256$ and $384$ nights of observations, with each night lasting $8$ hours. Further, for the third scenario, we present the results at redshift $7.1$ for $5120$ hours, assuming $640$ nights of observations. The increased observing time for the third scenario is due to the lower mean neutral hydrogen fraction outside the targeted bubble. 
Assuming all nights are the same, we reduce the system noise corresponding to an $8$ hrs observation by a factor of $16$, $19.6$, and $25.29$ for the three scenarios, respectively. The resulting mock signal becomes equivalent to $2048$, $3072$, and $5120$ hours of uGMRT observations. Similarly, for the SKA1-Low we present our results for $96$ hrs of observation for the first two cases at redshift $7.1$ and $8.3$ (neutral hydrogen fraction ${\sim 0.9}$) and for $200$ hrs of observation for the third case at redshift $7.1$ (neutral hydrogen fraction ${\sim 0.5}$). Eventually, we obtain mock visibility data with all the signal contributions (H{\footnotesize I} 21-cm signal, galactic foreground, extragalactic point sources, and system noise contribution) typical of the uGMRT and SKA1-Low observations.

\begin{figure}[htbp]
\centering
\includegraphics[width=0.49\textwidth]{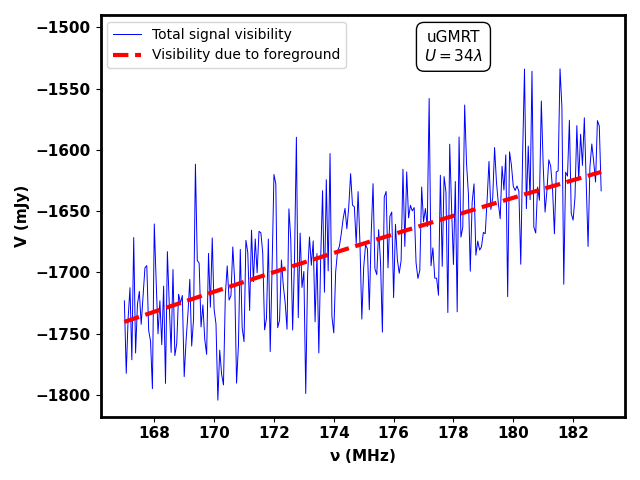}
\includegraphics[width=0.49\textwidth]{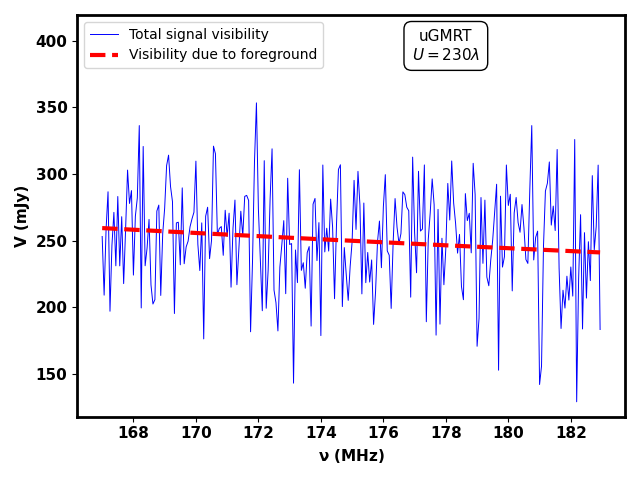}
\\
\includegraphics[width=0.49\textwidth]{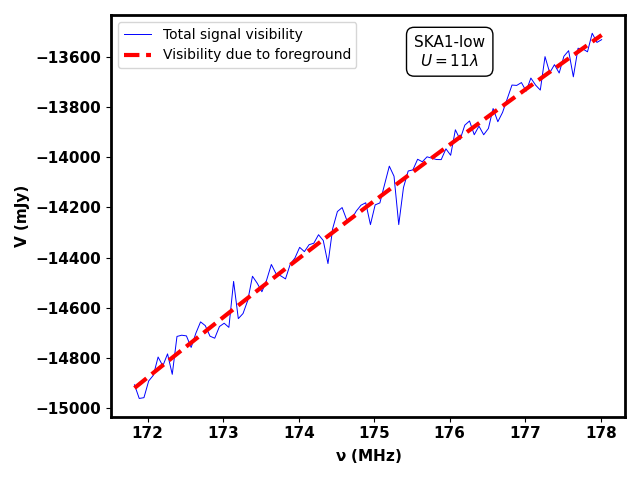}
\includegraphics[width=0.49\textwidth]{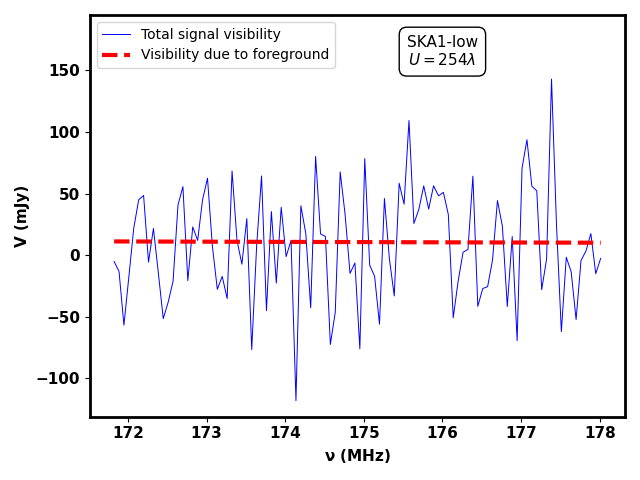}
\caption{Upper panels show mock visibilities (H{\footnotesize I} 21-cm signal + diffuse galactic foreground + system noise) as a function of observing frequency for the uGMRT at baselines $U= 34 \lambda$ (upper left panel) and $U=230 \lambda$ (upper right) at $175$ MHz. Lower panels show the same for the SKA1-Low at baselines $U= 11 \lambda$ (lower left panel) and $U=254 \lambda$ (lower right). Red dashed lines show diffuse galactic foreground contamination only.}
\label{fig:vis_tot}
\end{figure}

Figure \ref{fig:vis_tot} shows the mock visibility data comprising of H{\footnotesize I} signal, diffused galactic foreground contribution, and system noise as a function of frequency for two baselines $U=34\lambda$ (top left panel) and $U=230\lambda$ (top right panel) for the uGMRT at $175$ MHz. The corresponding mock visibility for the SKA1-Low is shown in the bottom panels (bottom left panel: $U=11\lambda$, and bottom right panel: $U=254\lambda$). They also show diffused galactic foreground contributions (red dashed lines). We see that the diffused galactic foreground dominates over the H{\footnotesize I} 21-cm signal and system noise at a small baseline (left panel) and, at a large baseline (right panel) the system noise dominates over the other two components. The H{\footnotesize I} 21-cm signal remains buried under the stronger foreground and system noise contributions.

\begin{figure}[htbp]
\centering
\includegraphics[width=0.49\textwidth]{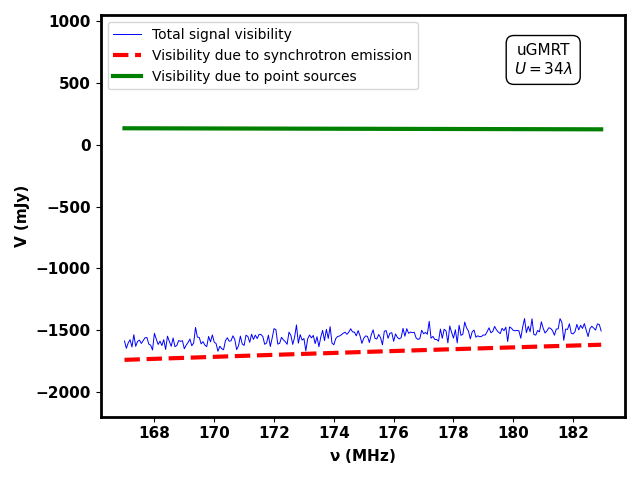}
\includegraphics[width=0.49\textwidth]{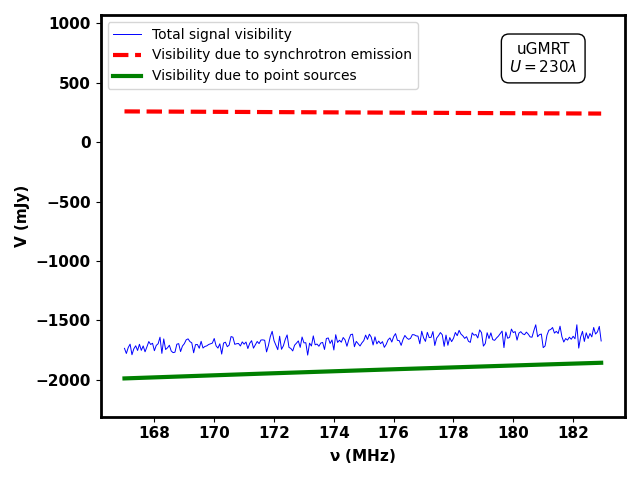}
\caption{Left panel shows the total mock visibilities (H{\footnotesize I} 21-cm signal + DGSE + radio point sources + system noise) as a function of observing frequency for the uGMRT at baseline ${U= 34 \lambda}$ and the right panel shows the same at baseline ${U=230 \lambda}$ at $175$ MHz observing frequency. Red dashed lines show diffuse galactic foreground contamination and the green lines show the point source contribution separately.}
\label{fig:vis_tot_ps_gmrt}
\end{figure}

Figure \ref{fig:vis_tot_ps_gmrt} shows the total mock visibility comprising of H{\footnotesize I} signal, diffused galactic synchrotron emission, radiation from extra-galactic radio point sources, and system noise as a function of frequency for two baselines ${U=34\lambda}$ (left panel) and ${U=230\lambda}$ (right panel) for the uGMRT at $175$ MHz. We note that Figure \ref{fig:vis_tot_ps_gmrt} shows contribution both from the extragalactic point sources, galactic diffuse synchrotron radiation along with other components unlike Fig. \ref{fig:vis_tot} which shows galactic diffuse synchrotron emission only as the foreground component. The radio point sources are modeled using a simplistic power-law form as discussed in subsection \ref{subsubsec:ps}. The green lines indicate visibilities due to point sources. We see that at a small baseline (left panel) the diffuse galactic synchrotron emission dominates over the point sources; at a large baseline (right panel), the point sources dominate over the diffuse galactic synchrotron emission. The total mock visibilities are similar for the second scenario at ${z=8.3}$ for both experiments, except for the fact that the system noise and foreground contributions are higher. Therefore, we do not show them here explicitly. The subsequent section discusses the subtraction of the foreground contribution.

Here, we would like to note that the visibility data used in this study does not contain calibration errors, ionospheric effects, and radio-frequency interference (RFI). A comprehensive investigation of these factors will be addressed in future work.

\section{Detecting individual ionized bubble in H{\footnotesize I} 21-cm maps}
\label{sec:ionized_bubble_detection}
\subsection{Foreground subtraction}

The foreground contributions are anticipated to be several orders of magnitude higher than the H{\footnotesize I} 21-cm signal around ionized bubbles during the EoR. Therefore, one significant challenge is reliably subtracting the foreground signal from the observed signal. All foreground mitigation techniques use the fact that the foregrounds are smooth along the frequency~\citep{Cho_2010, Martire_2022}. As discussed earlier, the foreground signal in our simulated mock data has also been assumed to be smooth along frequency. To mitigate foreground components, we fit the total mock visibility signal with a $3^{rd}$ order polynomial of observing frequency $\nu$ as follows,
\begin{equation}
    f(\nu)= a\nu^3+b\nu^2+c\nu+d, 
    \label{eq:polynomial}
\end{equation}
where parameters $a$, $b$, $c$, and $d$ are constant. Then, we find the best-fit polynomial along each line of sight corresponding to a baseline and subsequently subtract them from the total simulated mock visibility signal. \par

\begin{figure}[htbp]
\centering
    
\includegraphics[width=0.49\textwidth]{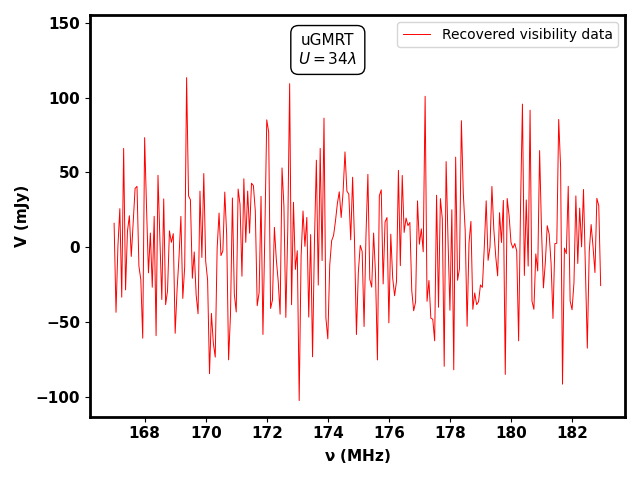}
\includegraphics[width=0.49\textwidth]{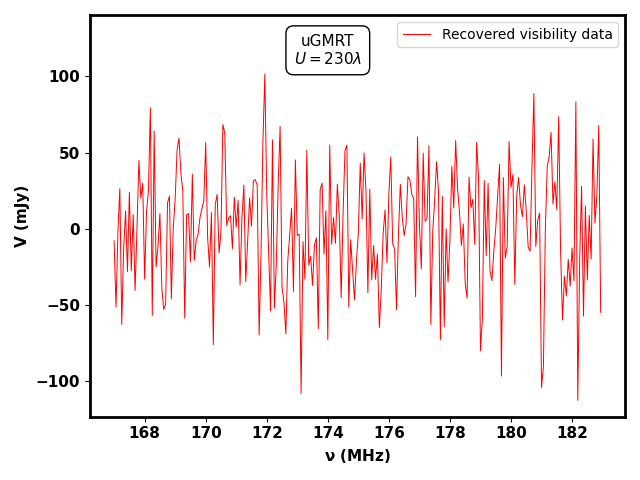}
\\
\includegraphics[width=0.49\textwidth]{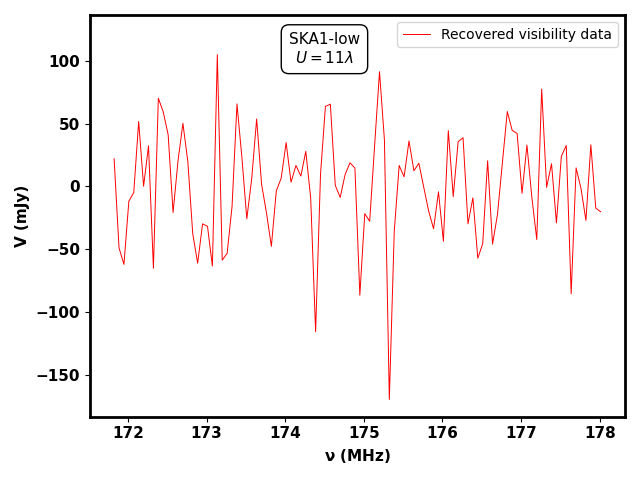}
\includegraphics[width=0.49\textwidth]{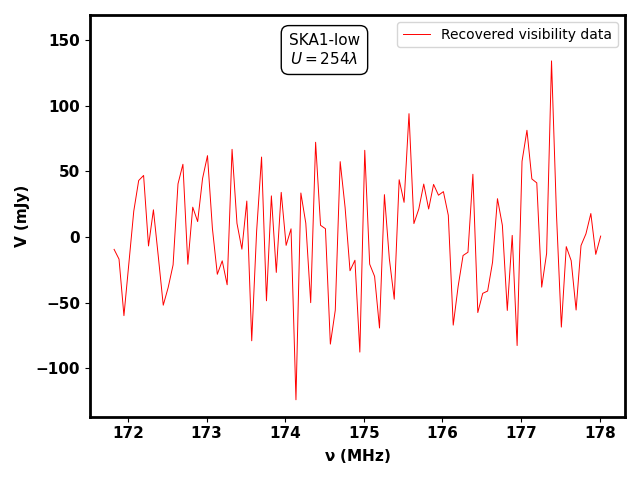}
        
\caption{This shows recovered visibilities after foreground subtraction for the uGMRT (top panels) and SKA1-Low (bottom panels) at the same baselines and frequency considered in Figure \ref{fig:vis_tot}.}
\label{fig:res_vis}

\end{figure}

\begin{figure}[htbp]
\centering
    
\includegraphics[width=0.49\textwidth]{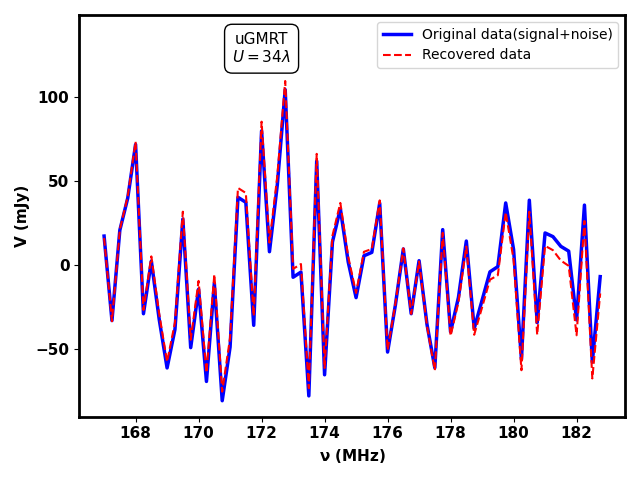}
\includegraphics[width=0.49\textwidth]{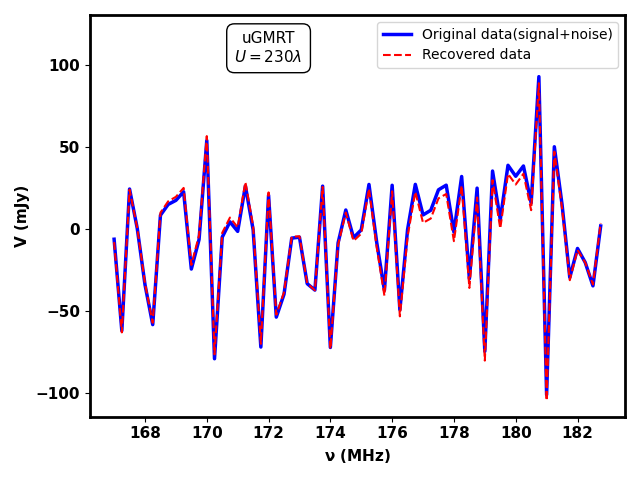}
\\
\includegraphics[width=0.49\textwidth]{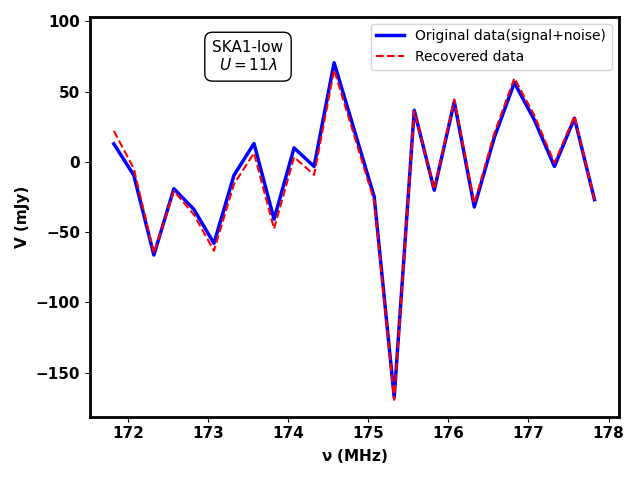}
\includegraphics[width=0.49\textwidth]{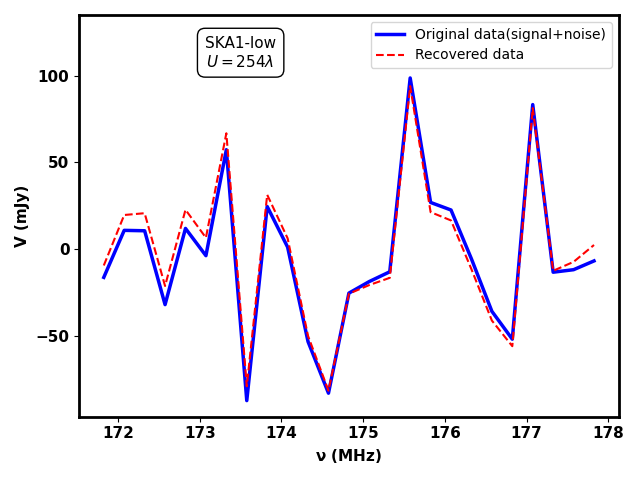}
        
\caption{These plots compare the recovered visibility (red dashed lines) \textbf{after subtracting the diffused galactic foreground contribution} with the original visibility consisting of the H{\footnotesize I} 21-cm signal and noise contributions (blue lines) for the uGMRT (top panels) and SKA1-Low (bottom panels) at same baselines and observing frequency considered in Figure \ref{fig:vis_tot}. {\bf Here, the foreground component consists of diffuse galactic radiation only and visibilities are plotted for every fourth frequency channel for visual clarity.}}
\label{fig:res_com}
\end{figure}

\begin{figure}[htbp]
\centering
    
\includegraphics[width=0.49\textwidth]{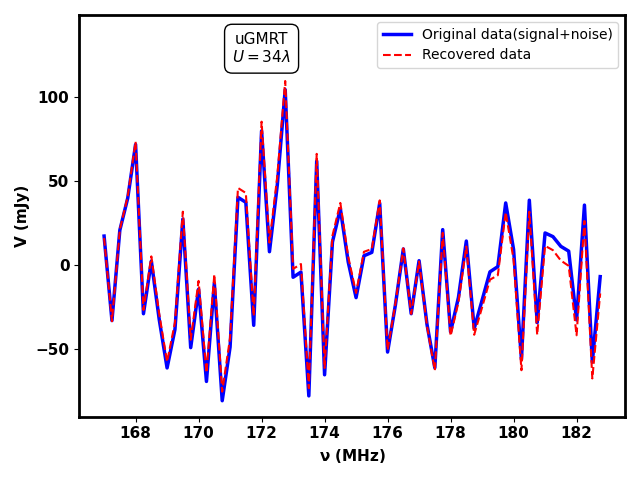}
\includegraphics[width=0.49\textwidth]{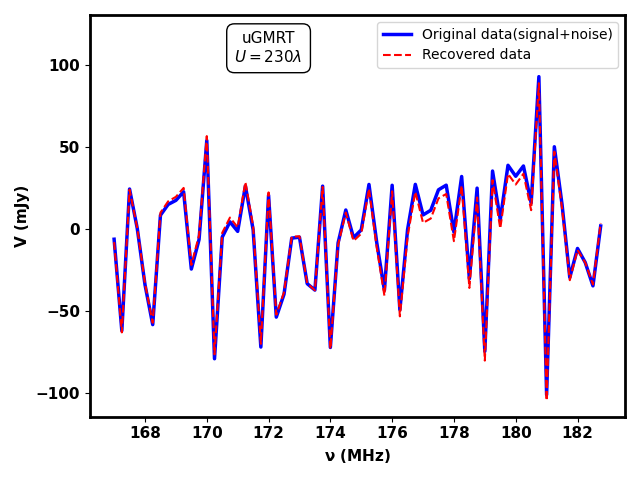}
        
\caption{\textbf{These plots compare the recovered visibility (red dashed lines) after subtracting the extra galactic point sources with the original visibility consisting of the H{\footnotesize I} 21-cm signal and noise contributions (blue lines) for the uGMRT at same baselines and observing frequency considered in the top panels of Figure \ref{fig:vis_tot}. Here, the foreground component consists of extra galactic point sources only. Visibilities are plotted for every fourth frequency channel for visual clarity.}}
\label{fig:res_com_ps}
\end{figure}

Figure \ref{fig:res_vis} presents the residual visibility at 175 MHz after the foreground contribution is subtracted from the total mock visibility for the uGMRT (top panels) and the SKA1-Low (bottom panels). Here, we have only considered the diffuse galactic synchrotron emission as the foreground contamination. \textbf{However, we get a very similar residual signal when we perform similar analysis on the total mock visibilities considering the radio point source contribution, which has been shown in Figure \ref{fig:res_com_ps}}. The residual signal contains the H{\footnotesize I} 21-cm signal, the system noise, and the un-subtracted foreground, if any. To assess the reliability of foreground subtraction, we compare the combined input H{\footnotesize I} 21-cm signal and system noise with the extracted signal for two different baselines ${U=34\lambda}$ (top left panel) and ${230\lambda}$ (top right panel), as depicted in Figure \ref{fig:res_com}. In the bottom panels of Figure \ref{fig:res_com}, we show a similar comparison for SKA1-Low for two different baselines ${U=11\lambda}$ (bottom left panel) and ${U=254\lambda}$ (bottom right panel). The plots show that the residual visibility closely matches the input visibility data, with only minor deviations. These slight deviations are possibly due to some residual foreground contribution and subtraction of the H{\footnotesize I} 21-cm signal and the system noise during the foreground subtraction. Consequently, we can conclude that we have mostly subtracted the foreground data from the mock visibility data, making it ready for further analysis, such as matched filtering. We do not present similar results for the second scenario at $z=8.3$ as it is similar to the first scenario.

\subsection{Matched filter}
The residual mock visibility (see Figure \ref{fig:res_vis}) contains mainly the H{\footnotesize I} 21-cm signal around the ionized bubble and the system noise from radio interferometers. We see that the dominance of system noise over the H{\footnotesize I} 21-cm signal is so pronounced that the latter's presence is barely discernible. Here, we employ a matched filter technique to maximize the signal-to-noise ratio, thereby enhancing the prospects of detecting the H{\footnotesize I} 21-cm signal around an ionized bubble. The functional form of the visibility signal corresponding to neutral hydrogen around a spherical ionized bubble is quite well known (refer to eq. \ref{eq:vis_HI}). Additionally, the system noise is anticipated to follow a Gaussian random distribution. These conditions enable us to implement the matched filter-based technique effectively. Below, we briefly summarize the method proposed in \cite{Datta_2007} and consider the following estimator for detecting ionized bubbles from the H{\footnotesize I} 21-cm signal, which is given as 
\begin{equation}
    {\hat{E}} = \left[{\sum_{a,b} {S_f}^* (\bm{U}_a, \nu_b) \hat{V}(\bm{U}_a, \nu_b)}\right]/\left[{\sum_{a,b}1}\right].
    \label{eq:estim}
\end{equation}
Here, $S_f(\bm{U}, \nu)$ is the filter which is a function of both baseline $\bm{U}$ and frequency $\nu$. $\hat{V}(\bm{U}_a, \nu_b)$ is the total observed visibility which consists of the desired H{\footnotesize I} 21-cm signal, residual foregrounds, system noise, etc. Here, $\bm{U}_a$ and $\nu_b$ refer to the different baselines and frequency channels in our observations. Here, we note that foreground components are subtracted from the total observed visibilities and we employ the matched filtering technique only on the residual visibility which contains only the H{\footnotesize I} 21-cm signal and system noise. Figure \ref{fig:baseline-dis} shows a typical uGMRT and SKA1-Low baseline distribution for $8$ hrs of observations. We note that the estimator is calculated by taking the sum over all baselines and frequency channels. We get one estimator for a given filter. The contribution of the system noise to the estimator is expected to be zero if it is averaged over a large number of independent realizations. However, for a single realization of the system noise, the contribution of it is unlikely to be exactly zero. It would change for different realizations of the system noise. Therefore, we estimate the variance of the estimator due to the system noise using the following equation \citep{Datta_2007}, 
\begin{equation}
   \left<(\Delta {\hat{E}})^2\right>_{NS}= {\langle N^2\rangle} \left[\sum_{a,b} \left| S_f(\bm{U}_a, \nu_b) \right|^2\right] / \left[\sum_{a,b} \right]^2.
\end{equation}
Here, $\langle N^2\rangle$ is the variance of the system noise for single visibility and can be calculated using eq. \ref{eq:sys-noise}. The symbol $\sum_{a,b}$ denotes summation over all baselines and frequency channels. We note that the variance to the estimator depends on the chosen filter, system noise, and baseline distribution but not on the desired signal. The signal-to-noise ratio (SNR) of a particular detection (or non-detection) can be calculated as,
\begin{equation}
    {\rm SNR}=\frac{\langle \hat{E} \rangle}{\sqrt{ \left<(\Delta {\hat{E}})^2\right>_{\rm NS}}}. 
    \label{eq:SNR}
\end{equation}
In this work, we consider ${\rm SNR} \gtrsim 5$ as the threshold for detection. We would like to mention that the SNR reaches its maximum when the selected filter exactly (or closely) matches with the H{\footnotesize I} 21-cm signal which is contaminated by the system noise. Therefore, the chosen filter has the same functional form as the H{\footnotesize I} 21-cm signal around an ionized bubble (refer to eq. \ref{eq:vis_HI}). It also has one free parameter i.e., the radius of the ionized bubble $R_{\rm b}$. We assume that the location of the bubble is known a priori from other observations. To find a filter that matches the targeted signal, one needs to explore the entire possible ranges of all parameters (here the size of the targeted bubble). The particular set of parameters yielding the maximum SNR will reveal the size of the ionized bubble in the observed H{\footnotesize I} 21-cm signal.

\section{Results}
\label{sec:results}
\subsection[Detectability of the QSO ionized bubble at \texorpdfstring{$z=7.1$}{z=7.1}]{Detectability of the QSO ionized bubble at $\bm{z=7.1}$}
\label{subsec:qso_7.1}
\begin{figure}[htbp]
\centering
\includegraphics[width=0.49\textwidth]{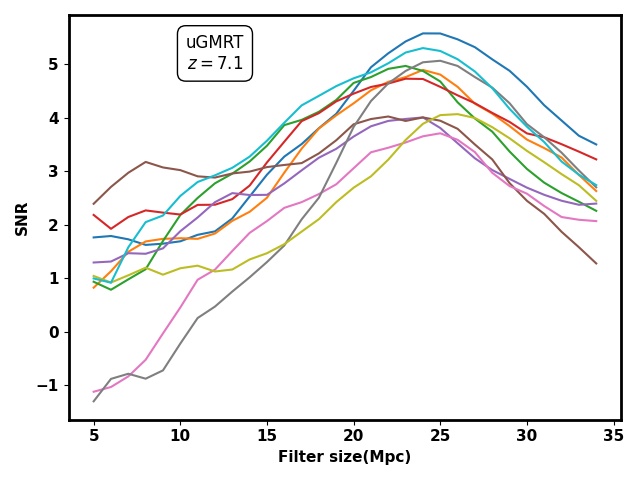}
\includegraphics[width=0.49\textwidth]{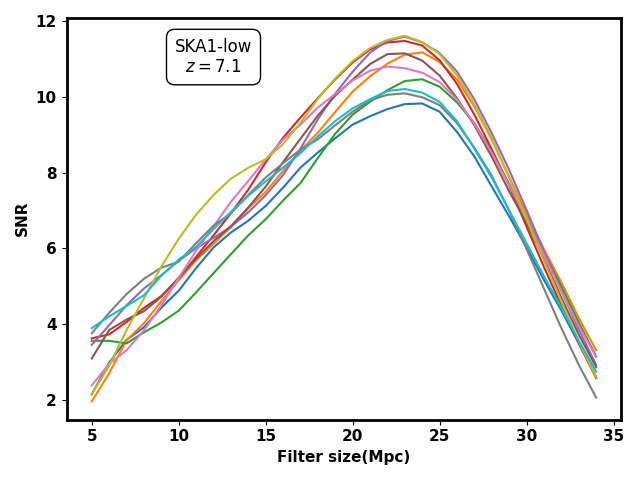}
\caption{This figure shows the signal-to-noise ratio for $10$ independent noise realizations corresponding to $2048$ hrs of observation time for the uGMRT (left panel) and $96$ hrs of SKA1-Low observations (right panel) at redshift $7.1$.} 
\label{fig:10_realizations}
\end{figure}

Here, we present our findings regarding the detectability of ionized bubbles around bright QSO/galaxies in H{\footnotesize I} 21-cm maps, employing the matched filter technique. We assume that the targeted ionized bubble is located at the antenna phase center. Thus, we simulate the targeted ionized bubble at the center of the simulation cube. Subsequently, we apply the matched filtering technique to the residual visibility obtained after subtracting foreground contaminants. As explained above the residual visibility contains the H{\footnotesize I} 21-cm signal, system noise, and unsubtracted foregrounds. First, we calculate the estimator ${\hat{E}}$ using equation \ref{eq:estim}. Because the actual size of the targeted ionized bubble is not known in real observations we calculate the estimator for various filter sizes ranging from $5$ to $35$ Mpc. Subsequently, we estimate the variance of the estimator $\left<(\Delta {\hat{E}})^2\right>_{\rm NS}$ due to the system noise using equation \ref{eq:sys-noise} and the signal to noise ratio (SNR, refer to eq. \ref{eq:SNR}) for the same range of filter sizes. Figure~\ref{fig:10_realizations} shows the resulting SNR as a function of filter size $R_{\rm f}$ for $2048$ hrs of uGMRT observation (left panel) and $96$ hrs of SKA1-Low observations (right panel) at $z=7.1$.

The ten different curves in Figure~\ref{fig:10_realizations} represent the SNR for ten independent realizations of the system noise. We see that for both the uGMRT and the SKA1-Low, the SNR peaks at filter size $\sim 23$ Mpc, with minor deviations in different noise realizations. As per the principle of the matched filter technique, the SNR peaks when the filter matches the actual targeted signal (i.e., the H{\footnotesize I} 21-cm signal around the ionized bubble) in the simulated H{\footnotesize I} maps. Therefore, we can conclude that the extracted radius of the ionized bubble in the simulated H{\footnotesize I} 21-cm map is around $\sim 23$ Mpc. We want to note that the actual radius of the ionized bubble measured directly from the simulated map is $23.5$ Mpc, which is very close to the extracted ones using the matched filter technique. The location of the peak in the SNR curve shifts around the mean value for different realizations, which we shall discuss in detail in the next section.\par
Further, we see that peak SNR values in all the ten realizations are $\gtrsim 4$ for $2048$ hrs of observations with the uGMRT and $\sim 10$ for $96$ hrs of observations with the SKA1-Low. We assume a total bandwidth of $16$ MHz and $6.25$ MHz while subtracting the foreground for the uGMRT and SKA1-Low, respectively. The choice of smaller bandwidth for the SKA1-Low is solely due to the limited computational resources available. Use of the higher bandwidth for the SKA1-Low would have subtracted the H{\footnotesize I} 21-cm signal less during the foreground subtraction process, thus increasing the SNR. Subsection \ref{fg-sub} discusses the impact of foreground subtraction in more detail. The higher SNR obtained in our results indicates that it is possible to significantly detect individual ionized bubbles in the H{\footnotesize I} 21-cm maps using the uGMRT with $2048$ hrs observations and using the SKA1-Low with only around $\sim 100$ hrs of observations.

\begin{figure}[htbp]
\centering
\includegraphics[width=0.49\textwidth] {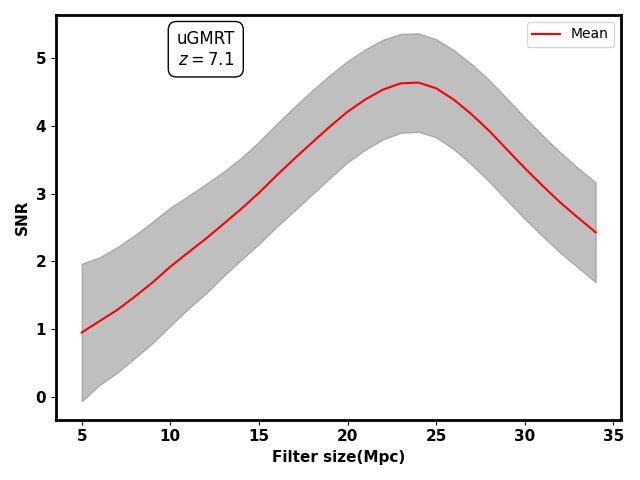}
\includegraphics[width=0.49\textwidth]{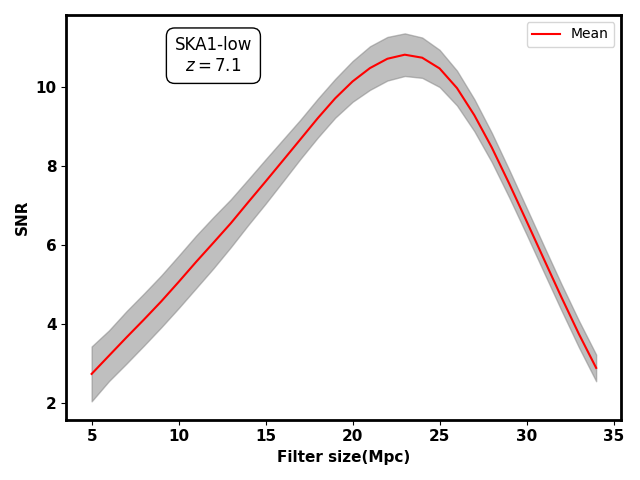}
\caption{The left panel shows the mean signal-to-noise (SNR) ratio of $100$ independent noise realizations as a function of filter size for $2048$ hrs of the uGMRT observations at redshift $7.1$. The right panel shows the SNR of $20$ independent noise realizations for $96$ hrs of the SKA1-Low observations. The shaded region shows $1\sigma$ uncertainty estimated from independent realizations.}
\label{fig:100_realizations}
\end{figure}

\begin{figure}[htbp]
\centering
\includegraphics[scale=.7]{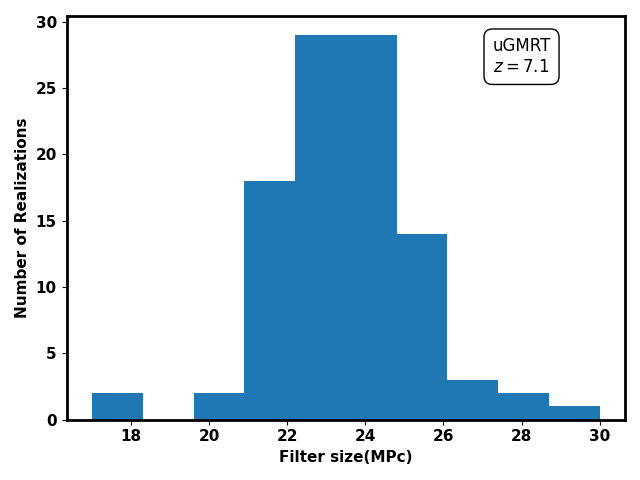}
\caption{This figure shows the histogram of filter sizes corresponding to peaks of the SNR for $100$ different noise realizations for the uGMRT at redshift $7.1$. }
\label{fig:hist}
\end{figure}

To assess the significance of the above result, we estimate the SNR for $100$ independent realizations of the system noise for the uGMRT and $20$ independent realizations for SKA1-Low, while keeping the H{\footnotesize I} 21-cm signal and foreground components the same. Figure \ref{fig:10_realizations} shows the SNR for some of these realizations for the uGMRT (left panel) and SKA1-Low (right panel) at $z=7.1$. The left panel of Figure \ref{fig:100_realizations} shows the corresponding mean SNR and $1-\sigma$ spread for all $100$ realizations of the uGMRT, while the right panel shows the same for the SKA1-Low with $20$ realizations. We see that all the realizations show a peak in the SNR. In principle, the peak should appear at the actual ionized bubble size of $23.5$ Mpc. However, the peak location varies in the range of $18-30$ Mpc and $22-24$ Mpc for the uGMRT and SKA1-Low, respectively, for different realizations. It can be seen in Figure~\ref{fig:hist}, which shows a histogram of peak locations for $100$ realizations for the uGMRT.

We see that, although in most cases, the peak is in the range of $22-26$ Mpc for the uGMRT, some extreme realizations show peaks at higher/lower filter radii. The peak value of the SNR for the uGMRT also changes in the range of $\sim 3-7$. Due to much lower system noise and higher sensitivity, the SKA1-Low should be able to detect the ionized bubble with reduced observing time and higher SNR. The estimated bubble size is also highly accurate for the SKA1-Low, which we do not explicitly show here. We note that the above analysis considers only the DGSE as the foreground contaminant. However, we have repeated the analysis including contributions from both the DGSE and radio point sources as foregrounds. We find very similar results to those presented here. Therefore, we do not explicitly show the outcomes of the combined foreground contamination here.

\subsection[Detectability of the ionized bubble at \texorpdfstring{$z=8.3$}{z=8.3}]{Detectability of the ionized bubble at $\bm{z=8.3}$}

\begin{figure}[htbp]
\centering
\includegraphics[width=0.49\textwidth]{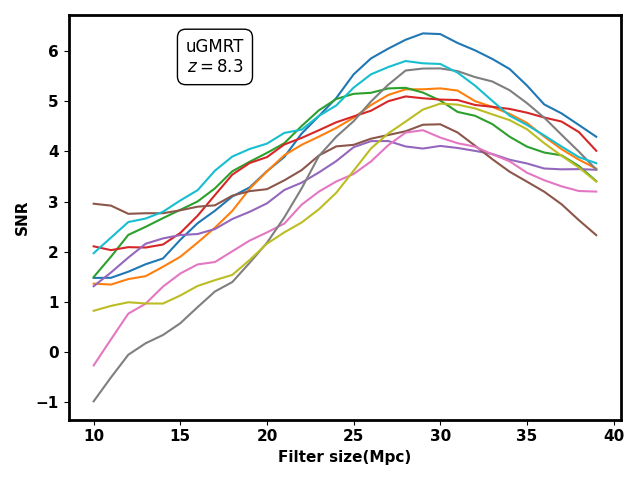}
\includegraphics[width=0.49\textwidth]{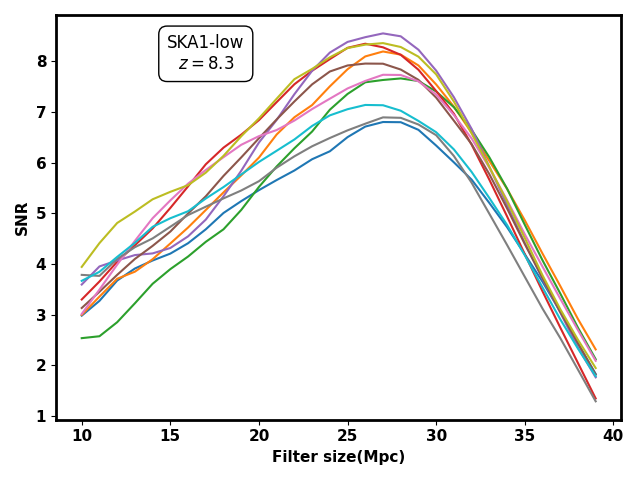}
\caption{This figure shows the signal-to-noise ratio for $10$ independent noise realizations corresponding to $3072$ hrs of observation time for the uGMRT (left panel) and $96$ hrs of SKA1-Low observations (right panel) at redshift $8.3$. } 
\label{fig:10_realizations_83}
\end{figure}

Unlike the ionized bubble at $z=7.1$, mainly created by a bright QSO, the bubble at $z=8.3$ is produced by a collection of galaxies as indicated in~\cite{witstok24}. This ionized bubble is relatively more aspherical in shape than at $z=7.1$. Figure~\ref{fig:10_realizations_83} presents the SNR as a function of filter size $R_{\rm f}$ for 3072 hrs of uGMRT (left panel) and $96$ hrs of SKA1-Low (right panel) observations at $z=8.3$.
Different curves are for different independent noise realizations. The system temperature is higher at $z=8.3$ compared to $z=7.1$. Therefore, the SNR decreases at a higher redshift for the same observation time. We see that for both the uGMRT and the SKA1-Low, the SNR peaks at a filter size $\sim 28$ Mpc, with minor deviations in different noise realizations. It is very close to the actual radius of the ionized bubble measured directly from the simulated map, which is $27.8$ Mpc. 

Further, we see that peak SNR values in all ten realizations are $\gtrsim 4$ for the uGMRT and $\sim 8$ for the SKA1-Low. Like the previous case, we assume a total bandwidth of $16$ MHz and $6.25$ MHz while subtracting the foreground for the uGMRT and SKA1-Low, respectively. We see that it is possible to reliably detect individual ionized bubbles in the H{\footnotesize I} 21-cm maps using the uGMRT with $3072$ hrs of observation time and the SKA1-Low with only $\sim 100$ hrs of observations at $z=8.3$.\par
\begin{figure}[htbp]
\centering
\includegraphics[width=0.49\textwidth]{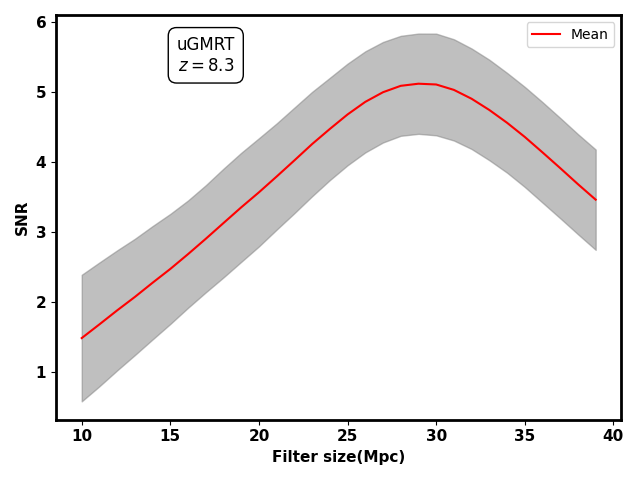}
\includegraphics[width=0.49\textwidth]{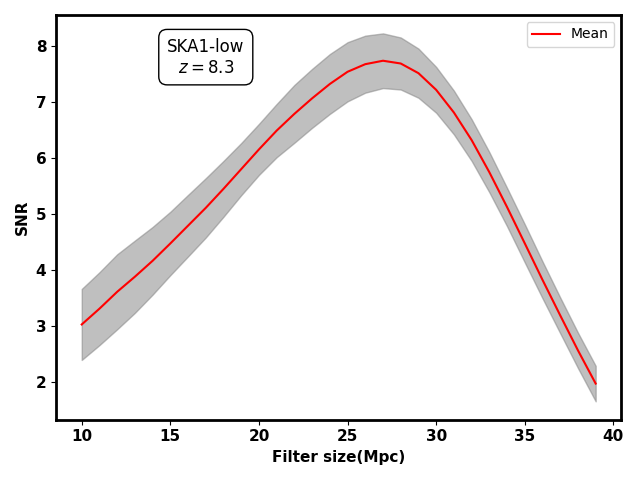}
\caption{The left panel shows the mean signal-to-noise (SNR) ratio of $100$ independent noise realizations as a function of filter size for $3072$ hrs of the uGMRT observations at redshift $8.3$. The right panel shows the SNR of $20$ independent noise realizations for $96$ hrs of the SKA1-Low observations. The shaded region shows $1\sigma$ uncertainty estimated from independent realizations.}
\label{mean_83}
\end{figure}
\begin{figure}[htbp]
\centering
\includegraphics[scale=.7]{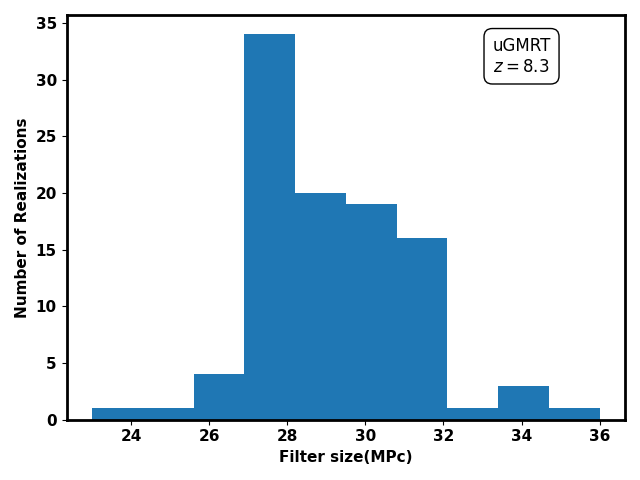}
\caption{ Same as Figure \ref{fig:hist} at redshift $z=8.3$.}
\label{fig:hist_83}
\end{figure}

The left panel of Figure~\ref{mean_83} shows the mean SNR along with the $1-\sigma$ spread for $100$ noise realizations of the uGMRT, while the right panel shows the same for the SKA1-Low with $20$ realizations at $z=8.3$. Further, the peak location varies between $24$ and $36$ Mpc for the uGMRT for different realizations (Figure~\ref{fig:10_realizations_83}), which can be seen in Figure~\ref{fig:hist_83}, which shows a histogram of peak locations for $100$ realizations for the uGMRT. The peak value of the SNR for the uGMRT also changes in the range of $\sim 4-6$ and $6.5-8.5$ for the SKA1-Low.

\subsection[Detectability of the QSO ionized bubble at \texorpdfstring{$z=7.1$}{z=7.1} with \texorpdfstring{$x_{\rm HI}=0.52$}{x{\rm HI}=0.52}]{Detectability of the QSO ionized bubble at $\bm{z=7.1}$ for $\bm{x_{\rm HI}=0.52}$}
\label{subsec:qso_7.1_0.52}

\begin{figure}[htbp]
\centering
\includegraphics[width=0.49\textwidth]{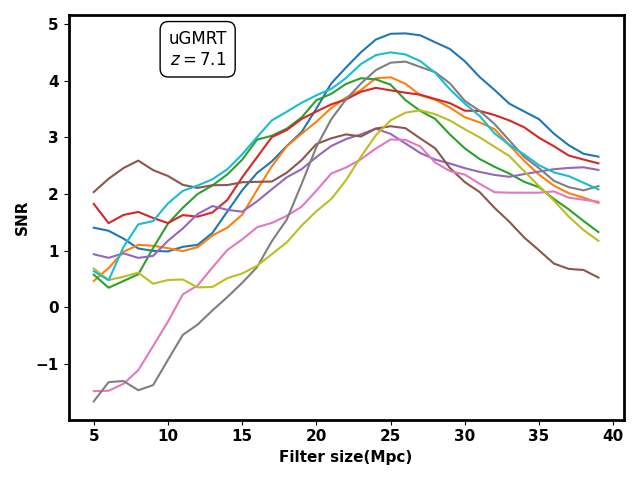}
\includegraphics[width=0.49\textwidth]{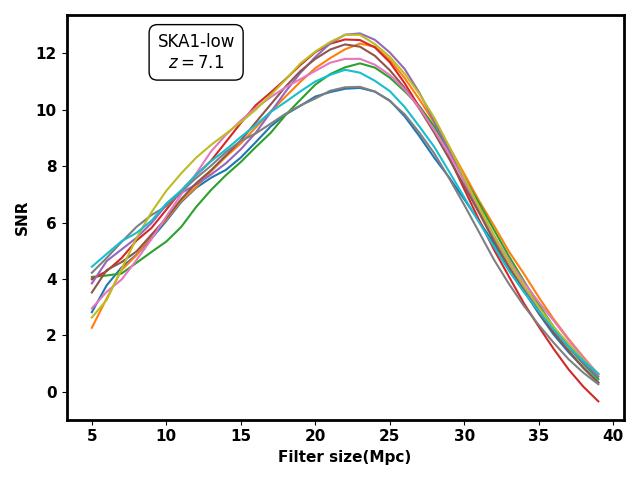}
\caption{This figure shows the signal-to-noise ratio for $10$ independent noise realizations corresponding to $5120$ hours of observation time for the uGMRT (left panel) and $200$ hours of SKA1-Low observations (right panel) at redshift $7.1$.}  
\label{fig:10_realizations_71_52}
\end{figure}

Here, we present our findings for the third case, where the targeted ionized bubble is surrounded by other ionized regions which result in a lower neutral hydrogen fraction of ${x_{\rm HI}=0.52}$. Further, the targeted bubble at the center also became aspherical. A lower neutral hydrogen fraction means that the background H{\footnotesize I} 21-cm signal is weaker, which results in a lower SNR value. We use the same matched filtering technique as described earlier but increased the observation time to $5120$ hours for the uGMRT and $200$ hours for the SKA1-Low, considering the lower SNR and increased difficulty in detecting ionized regions at a lower neutral fraction. Figure \ref{fig:10_realizations_71_52} shows the SNR as a function of filter size ${R_{\rm f}}$ for $5120$ hours of uGMRT observation (left panel) and $200$ hours of SKA1-Low observations (right panel) at $z=7.1$. As in previous cases, each curve represents a different noise realization. We see that for the uGMRT, the SNR peaks at a filter size ${\sim 25}$ Mpc, and for the SKA1-Low, it peaks at ${\sim 23}$ Mpc, with minor deviations in different noise realizations.

Further, we see that peak SNR values in all ten realizations are ${\gtrsim 3}$ for the uGMRT and ${\sim 11}$ for the SKA1-Low. Like the previous cases, we assume a total bandwidth of $16$ MHz and $6.25$ MHz while subtracting the foreground for the uGMRT and SKA1-Low, respectively. We see that it is possible to reliably detect individual ionized bubbles in the H{\footnotesize I} 21-cm maps using the uGMRT with $5120$ hours of observation time and the SKA1-Low with only $\sim 200$ hours of observations at $z=7.1$ with significantly much lower neutral hydrogen fraction value.

\begin{figure}[ht]
\centering
\includegraphics[width=0.49\textwidth]{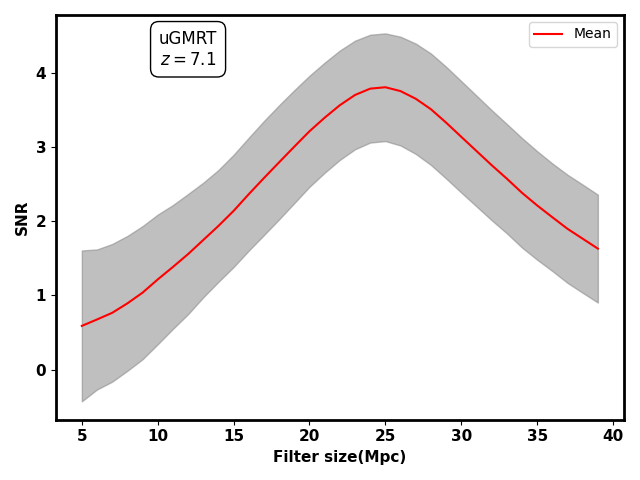}
\includegraphics[width=0.49\textwidth]{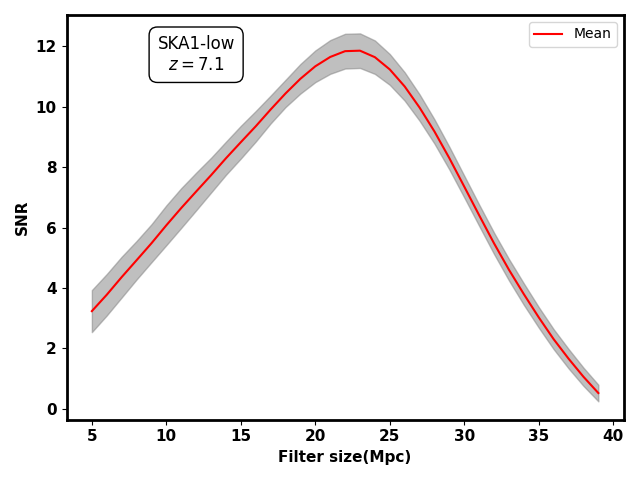}
\caption{The left panel shows the mean signal-to-noise (SNR) ratio of $100$ independent noise realizations as a function of filter size for $5120$ hours of the uGMRT observations at redshift $7.1$. The right panel shows the SNR of $20$ independent noise realizations for $200$ hours of the SKA1-Low observations. The shaded region shows ${1\sigma}$ uncertainty estimated from independent realizations.}
\label{mean_71_52}
\end{figure}

\begin{figure}[ht]
\centering
\includegraphics[scale=.7]{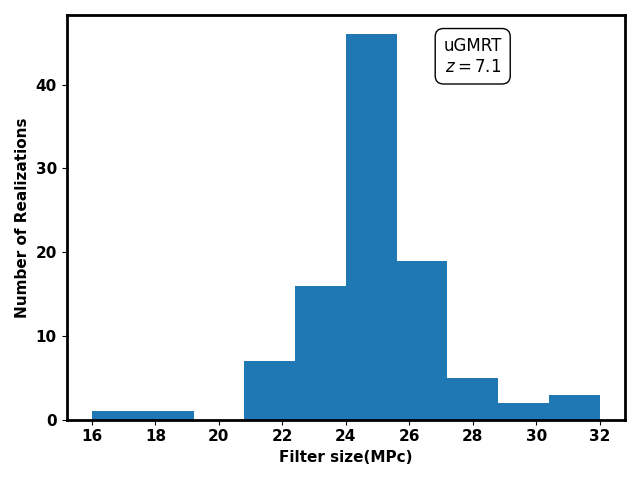}
\caption{Same as Figure \ref{fig:hist} at redshift $z=7.1$ with a lower neutral hydrogen fraction of $ 0.52$.}
\label{fig:hist_71_52}
\end{figure}

The left panel of Figure~\ref{mean_71_52} shows the mean SNR along with the ${1-\sigma}$ spread for $100$ noise realizations of the uGMRT, while the right panel shows the same for the SKA1-Low with $20$ realizations at $z=7.1$ for ${x_{\rm HI}=0.52}$. Further, the peak location varies between $16$ and $32$ Mpc for the uGMRT for different realizations (Figure~\ref{fig:10_realizations_71_52}), which can be seen in Figure~\ref{fig:hist_71_52}, that shows a histogram of peak locations for $100$ realizations for the uGMRT. The peak value of the SNR for the uGMRT also changes in the range of $\sim 3-5$ and $10.5-12.5$ for the SKA1-Low.

\subsection{Impact of foreground subtraction}
\label{fg-sub}

 \begin{figure}[htbp]
\centering
\includegraphics[scale=.7]{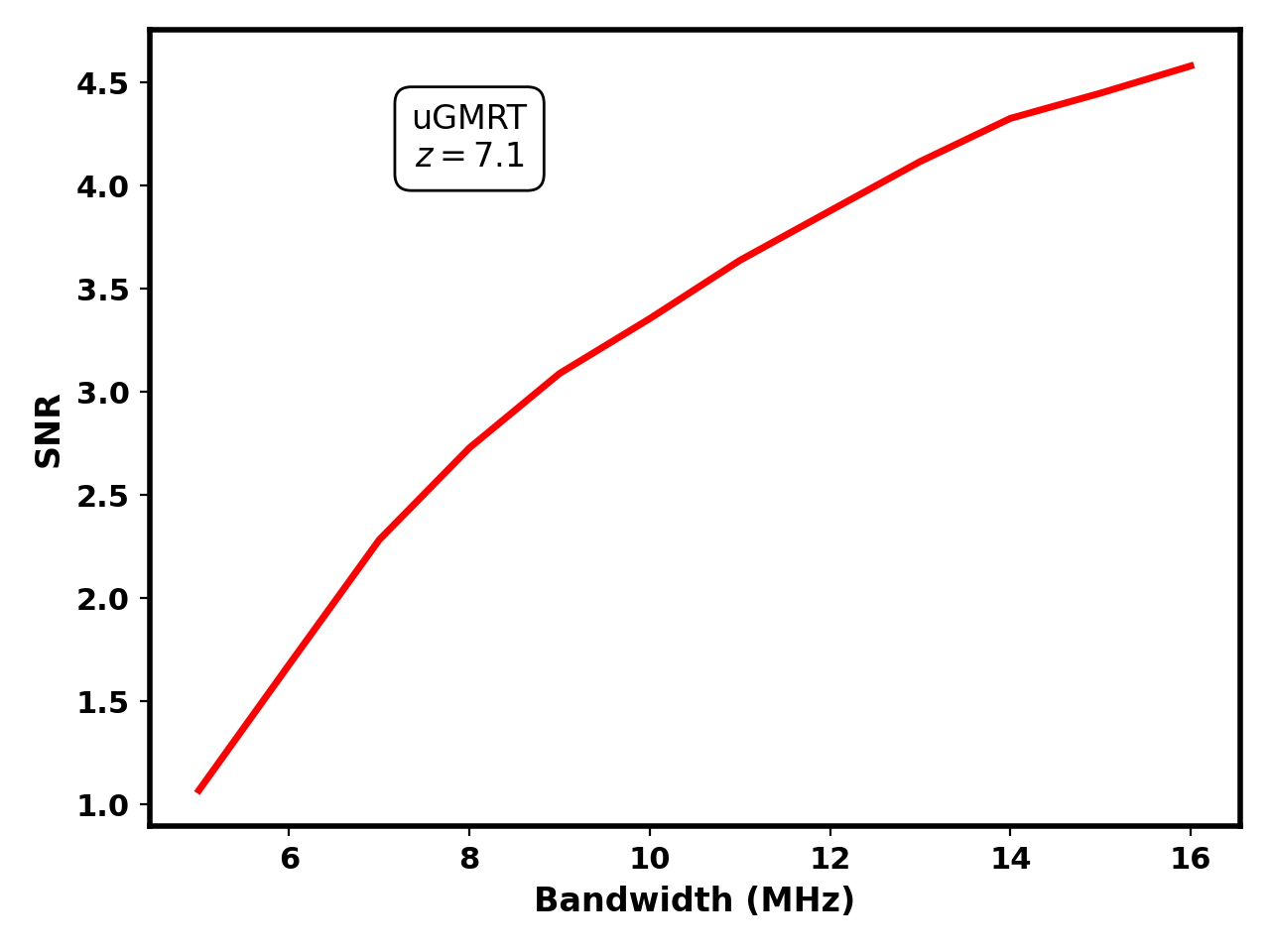}
\caption{This shows the signal-to-noise ratio as a function of frequency bandwidth used during foreground subtraction for the uGMRT at redshift $z=7.1$.} 
\label{fig:SNRvsBW}
\end{figure}

We noticed that during the process of foreground subtraction using a polynomial along the frequency axis, a small part of the H{\footnotesize I} 21-cm signal also gets subtracted along with the foreground. It reduces the SNR to some extent. We further notice that the amount of subtraction of the H{\footnotesize I} 21-cm signal depends on the frequency bandwidth used during the foreground subtraction process. The fraction of H{\footnotesize I} 21-cm signal subtracted is more for smaller bandwidth compared to the case when larger frequency bandwidth is used. Consequently, the drop in the SNR is more significant for smaller bandwidths. Figure~\ref{fig:SNRvsBW} shows the SNR as a function of the bandwidth for $2048$ hrs of observations with uGMRT at $z=7.1$. We see that the SNR, which is lower than $2$ at $6$ MHz bandwidth, increases to $4.5$ when $16$ MHz bandwidth is assumed. Here, we note that the SNR reaches $6.5$ for the above observation when no foreground is used and the process of foreground subtraction is skipped. Therefore, the SNR is reduced by $30$ percent when the bubble size is $\sim 23.5$ Mpc, and $16$ MHz bandwidth is assumed for foreground subtraction. Using higher bandwidth during the foreground subtraction would enhance the SNR and prospects of ionized bubble detection. However, increased bandwidth will increase the computational cost during the analysis. We find similar results for the SKA1-Low but do not explicitly show them here. The impact of foreground subtraction on the SNR will also depend on the foreground subtraction methods, and therefore, a thorough investigation is required.

\subsection{Constraining the neutral hydrogen fraction}

We see that the visibility corresponding to the H{\footnotesize I} 21-cm signal around a spherical ionized bubble in the IGM is proportional to the mass-averaged neutral hydrogen fraction $x_{\rm HI}$ of the IGM outside the bubble (refer to eq. \ref{eq:vis_HI}). One can further show that \citep{Datta_2009}, 
\begin{equation}
 {\rm SNR}=  C x_{\rm HI}, 
 \label{eq:snr}
\end{equation}
where $C$ depends on the bubble size, redshift, and various observational parameters such as the $T_{\rm sys}$, total number of antennae, observing time, antenna collecting area, and baseline distribution. All these quantities are constant at a fixed redshift. One can use this relation to place constraints on $x_{\rm HI}$. Now, we focus on the ionized bubble around the QSO at $z=7.1$ and $2048$ hrs of observation time with uGMRT. The SNR value changes for different noise realizations in Figure \ref{fig:10_realizations} (left panel). The average SNR value peaks at $4.6$ (Figure~\ref{fig:100_realizations}) at the filter size 23 Mpc for the uGMRT at $z=7.1$. Further, we see that $1\sigma$ uncertainty in the SNR value at the same filter size is 0.73. Therefore, we estimate the hydrogen neutral fraction using $x_{\rm HI} = ({\rm SNR} \pm 1\sigma_{\rm SNR})/C$, where the SNR and $1\sigma_{\rm SNR}$ is measured at the peak location.\par 
To calculate $C$, we simulate an entirely neutral and uniform H{\footnotesize I} 21-cm signal around a spherical ionized bubble of radius 23 Mpc, i.e., $x_{\rm HI}=1$ outside the ionized bubble. It allows us to estimate $C$ using eq.~\ref{eq:snr}. Now, after adding the foreground component only with the H{\footnotesize I} 21-cm signal, we follow the entire analysis and calculate the SNR as a function of filter size without adding system noise from the radio interferometers. The SNR, in this case, peaks at 23 Mpc and is supposed to be equal to $C$ (following eq.~\ref{eq:snr}) as the $x_{\rm HI}=1$ outside the ionized bubble in this case. We use this value of $C$ to estimate the mean $x_{\rm HI}$ and the $1\sigma$ uncertainty associated with it. The estimated neutral fraction is $x_{\rm HI}=0.93 \pm 0.15$ at $z=7.1$ for $2048$ hrs of observations with uGMRT. It is very close to the actual $x_{\rm HI}$ in the simulation, which is $0.88$. The estimated $1\sigma$ upper limit of $x_{\rm HI}$ exceeds unity which is, of course, unrealistic. The SKA1-Low places much tighter constraints, which is $x_{\rm HI}=0.92 \pm 0.055$ using only $96$ hrs of observations for the same scenario. Similarly we also studied this for the third case, where the neutral hydrogen fraction is lower in the simulation, discussed in subsection \ref{subsec:qso_7.1_0.52}. We find the estimated neutral hydrogen fraction to be ${x_{\rm HI}=0.49 \pm 0.09}$ at $z=7.1$ for $5120$ hours of observations with the uGMRT, which is very close to the actual neutral hydrogen fraction ${x_{\rm HI}=0.52}$.\par

The above result demonstrates the prospects of constraining the mass-averaged neutral fraction using the technique discussed in this work. The SKA1-Low is expected to put much tighter constraints on the $x_{\rm HI}$. A thorough analysis is required considering different bubble sizes and hydrogen neutral fractions at different redshifts for various instruments, and we leave this exercise for future work.

\section{Summary \& Discussion}
\label{sec:summary}
Several bright QSOs/galaxies have recently been detected at $z \gtrsim 6$. We expect large ionized bubbles, with comoving sizes of tens of Mpc, around these sources, which are embedded in a neutral H{\footnotesize I} medium. We have conducted a detailed investigation into the prospects of detecting such individual ionized bubbles using observations of the H{\footnotesize I} 21-cm signal from the epoch of reionization by the uGMRT and SKA1-Low. We have mainly focused on two known sources: one very bright QSO at $z=7.1$ \citep{Mortlock_2011} and a large ionized bubble around a collection of galaxies at redshift $8.3$ \citep{witstok24}. 

First, we simulated the H{\footnotesize I} 21-cm signal around bright QSOs/galaxies matching the above cases and foreground contaminants in a 3D cube. Subsequently, we have generated visibilities corresponding to the H{\footnotesize I} 21-cm signal and foregrounds at baselines similar to the uGMRT and SKA1-Low experiments and added the system noise with the visibilities corresponding to the H{\footnotesize I} signal and foreground contaminants. We then employ a foreground mitigation technique based on a polynomial fitting algorithm to remove them from the total visibilities. The residual visibilities match well with the combination of the input H{\footnotesize I} signal and system noise. However, it is heavily dominated by the system noise. To minimize the effects of system noise in the estimated quantity, we employed a technique based on the matched filter algorithm. This technique is applicable in our case as the functional form of the desired H{\footnotesize I} 21-cm signal around bright sources is known.  We find that the expected signal-to-noise ratio for the proposed estimator for the uGMRT is ${\sim 5}$ for both ${\sim 2000}$ hours of observations at redshift $7.1$ and ${\sim 3000}$ hours at redshift $8.3$, when the mean H{\footnotesize I} fraction outside the targeted bubble is ${\sim 0.9}$. However, the required observing time increases to ${\sim 5120}$ hrs for the uGMRT at redshift $7.1$ to achieve a similar SNR when the mean H{\footnotesize I} fraction is lower at ${\sim 0.5}$. The SKA1-Low will have a much higher SNR ($\sim 8 - 10$) with just $\sim 100$ hrs of observations at same redshifts $7.1$ and $8.3$, when the mean H{\footnotesize I} fraction outside the targeted bubble is ${\sim 0.9}$. However, the required observing time increases to ${\sim 200}$ hrs for the SKA1-Low at redshift $7.1$ to achieve a similar SNR when the mean H{\footnotesize I} fraction is lower at ${\sim 0.5}$.

Apart from detecting the H{\footnotesize I} 21-cm signal, the technique is also able to measure the size of the ionized regions and H{\footnotesize I} neutral fraction. For the uGMRT $\sim 2000$ hours of observations at redshift $7.1$, we find that the estimated bubble size ranges between $22-26$ Mpc for $100$ independent noise realizations when the actual bubble size is $23.5$ Mpc. Similar results are obtained for ${\sim 5120}$ hours of observations with the uGMRT at redshift $7.1$ with a mean H{\footnotesize I} fraction of $\sim 0.5$ and for ${\sim 3000}$ hours at redshift $8.3$ with a mean H{\footnotesize I} fraction of ${\sim 0.9}$. The SKA1-Low should measure the bubble size more accurately, with the uncertainty in bubble size estimation being less than $1$ Mpc at both mentioned redshifts. 

We have also explored the potential to constrain the neutral hydrogen fraction of the IGM outside the ionized bubble. We find that our technique can either directly measure the neutral hydrogen fraction or put an upper or lower limit on it, depending on the actual neutral fraction and the observational specifications.

The process of foreground subtraction using a polynomial along the frequency axis also removes some of the desired H{\footnotesize I} signal. We found that the fraction of subtracted H{\footnotesize I} signal increases when using smaller frequency bandwidth, significantly reducing the signal-to-noise ratio. We thoroughly investigated this effect using a wide range of bandwidths and found that the SNR increases by more than four times when the frequency bandwidth is increased from $5$ MHz to $16$ MHz for bubble size $23.5$ Mpc. We studied three particular cases here and results may vary if sizes of ionized bubbles and redshifts change. However, our methods can be applied to any similar cases.

Here, we focused on the targeted detection of ionized bubbles around sources previously detected by other telescopes. Several such sources have been observed during the reionization epoch. Additionally, the matched filter technique can detect ionized bubbles even without prior source detection. However, this unguided detection requires exploring a vast parameter space, which is computationally expensive. The underlying bubble may not be spherical, as assumed in our study. However, as long as the bubble is relatively isolated, the method remains applicable, though the SNR might be reduced. \textbf{Furthermore, we note that our simulations do not account for the contributions of low-mass halos to reionization. This omission likely results in an underestimation of the diffuse and complex structure of the ionized regions. Accurately incorporating these contributions could, to some extent, degrade the predicted signal-to-noise ratio.} Foregrounds are assumed to be spectrally smooth here, which may not be accurate in actual observations. A more sophisticated method for foreground mitigation may be needed in case foregrounds behave differently. We would also like to note that this study does not account for potential errors arising from improper calibration \citep{saikat24}, ionospheric effects, or radio-frequency interference (RFI). These practical issues will be addressed in future works.\\

\acknowledgments
AM and CSM acknowledge financial support from Council of Scientific and Industrial Research (CSIR) via  CSIR-SRF fellowships under grant no. 09/0096(13611)/2022-EMR-I and 09/1022(0080)/2019-EMR-I respectively. KKD acknowledges financial support from SERB (Govt. of India) through a project under MATRICS scheme (MTR/2021/000384). SM acknowledges financial support through the project titled “Observing the Cosmic Dawn in Multicolour using Next Generation Telescopes” funded by the Science and Engineering Research Board (SERB), Department of Science and Technology, Govt. of India through the Core Research Grant No. CRG/2021/004025. The authors acknowledge the use of computational resources available to the Inter-University Centre for Astronomy and Astrophysics (IUCAA), Pune, and to the Cosmology with Statistical Inference (CSI) research group at the Indian Institute of Technology Indore (IIT Indore).

\bibliographystyle{JHEP}
\bibliography{biblio}

\providecommand{\href}[2]{#2}\begingroup\raggedright\begin{thebibliography}{10}

\bibitem{Barkana_2001}
R.~Barkana and A.~Loeb, \emph{In the beginning: the first sources of light and the reionization of the universe}, \href{https://doi.org/10.1016/s0370-1573(01)00019-9}{\emph{Physics Reports} {\bfseries 349} (2001) 125}.

\bibitem{Loeb+2010+i+vi}
A.~Loeb, \emph{Frontmatter},  in \emph{How Did the First Stars and Galaxies Form?}, (Princeton), pp.~i--vi, Princeton University Press (2010), \href{https://doi.org/doi:10.1515/9781400834068.fm}{DOI}.

\bibitem{Pritchard_2012}
J.R.~Pritchard and A.~Loeb, \emph{21 cm cosmology in the 21st century}, \href{https://doi.org/10.1088/0034-4885/75/8/086901}{\emph{Reports on Progress in Physics} {\bfseries 75} (2012) 086901}.

\bibitem{bera2023}
A.~{Bera}, R.~{Ghara}, A.~{Chatterjee}, K.K.~{Datta} and S.~{Samui}, \emph{{Studying cosmic dawn using redshifted HI 21-cm signal: A brief review}}, \href{https://doi.org/10.1007/s12036-022-09904-w}{\emph{Journal of Astrophysics and Astronomy} {\bfseries 44} (2023) 10} [\href{https://arxiv.org/abs/2210.12164}{{\ttfamily 2210.12164}}].

\bibitem{mellema2006}
G.~{Mellema}, I.T.~{Iliev}, U.-L.~{Pen} and P.R.~{Shapiro}, \emph{{Simulating cosmic reionization at large scales - II. The 21-cm emission features and statistical signals}}, \href{https://doi.org/10.1111/j.1365-2966.2006.10919.x}{\emph{\mnras} {\bfseries 372} (2006) 679} [\href{https://arxiv.org/abs/astro-ph/0603518}{{\ttfamily astro-ph/0603518}}].

\bibitem{Furlanetto_2006}
S.R.~Furlanetto, S.P.~Oh and F.H.~Briggs, \emph{Cosmology at low frequencies: The 21cm transition and the high-redshift universe}, \href{https://doi.org/10.1016/j.physrep.2006.08.002}{\emph{Physics Reports} {\bfseries 433} (2006) 181}.

\bibitem{tirth2009}
T.R.~{Choudhury}, M.G.~{Haehnelt} and J.~{Regan}, \emph{{Inside-out or outside-in: the topology of reionization in the photon-starved regime suggested by Ly{\ensuremath{\alpha}} forest data}}, \href{https://doi.org/10.1111/j.1365-2966.2008.14383.x}{\emph{\mnras} {\bfseries 394} (2009) 960} [\href{https://arxiv.org/abs/0806.1524}{{\ttfamily 0806.1524}}].

\bibitem{Mortlock_2011}
D.J.~Mortlock, S.J.~Warren, B.P.~Venemans, M.~Patel, P.C.~Hewett, R.G.~McMahon et~al., \emph{A luminous quasar at a redshift of z = 7.085}, \href{https://doi.org/10.1038/nature10159}{\emph{Nature} {\bfseries 474} (2011) 616–619}.

\bibitem{Wu_2015}
X.-B.~Wu, X.-B.~Wu, F.~Wang, F.~Wang, F.~Wang, X.~Fan et~al., \emph{An ultraluminous quasar with a twelve-billion-solar-mass black hole at redshift 6.30}, \href{https://doi.org/10.1038/nature14241}{\emph{Nature} (2015) }.

\bibitem{Ba_ados_2017}
E.~Bañados, B.P.~Venemans, C.~Mazzucchelli, E.P.~Farina, F.~Walter, F.~Wang et~al., \emph{An 800-million-solar-mass black hole in a significantly neutral universe at a redshift of 7.5}, \href{https://doi.org/10.1038/nature25180}{\emph{Nature} {\bfseries 553} (2017) 473–476}.

\bibitem{Wang_2018}
F.~Wang, J.~Yang, X.~Fan, M.~Yue, X.-B.~Wu, J.-T.~Schindler et~al., \emph{The discovery of a luminous broad absorption line quasar at a redshift of 7.02}, \href{https://doi.org/10.3847/2041-8213/aaf1d2}{\emph{The Astrophysical Journal Letters} {\bfseries 869} (2018) L9}.

\bibitem{Wang_2019}
F.~{Wang}, J.~{Yang}, X.~{Fan}, X.-B.~{Wu}, M.~{Yue}, J.-T.~{Li} et~al., \emph{{Exploring Reionization-era Quasars. III. Discovery of 16 Quasars at 6.4 {\ensuremath{\lesssim}} z {\ensuremath{\lesssim}} 6.9 with DESI Legacy Imaging Surveys and the UKIRT Hemisphere Survey and Quasar Luminosity Function at z {\ensuremath{\sim}} 6.7}}, \href{https://doi.org/10.3847/1538-4357/ab2be5}{\emph{\apj} {\bfseries 884} (2019) 30} [\href{https://arxiv.org/abs/1810.11926}{{\ttfamily 1810.11926}}].

\bibitem{Matsuoka_2019a}
Y.~{Matsuoka}, M.~{Onoue}, N.~{Kashikawa}, M.A.~{Strauss}, K.~{Iwasawa}, C.-H.~{Lee} et~al., \emph{{Discovery of the First Low-luminosity Quasar at z > 7}}, \href{https://doi.org/10.3847/2041-8213/ab0216}{\emph{\apjl} {\bfseries 872} (2019) L2} [\href{https://arxiv.org/abs/1901.10487}{{\ttfamily 1901.10487}}].

\bibitem{Yang_2020}
J.~Yang, F.~Wang, X.~Fan, J.F.~Hennawi, F.B.~Davies, M.~Yue et~al., \emph{Pōniuā‘ena: A luminous z = 7.5 quasar hosting a 1.5 billion solar mass black hole}, \href{https://doi.org/10.3847/2041-8213/ab9c26}{\emph{The Astrophysical Journal Letters} {\bfseries 897} (2020) L14}.

\bibitem{Wang_2021}
F.~Wang, J.~Yang, X.~Fan, J.F.~Hennawi, A.J.~Barth, E.~Banados et~al., \emph{A luminous quasar at redshift 7.642}, \href{https://doi.org/10.3847/2041-8213/abd8c6}{\emph{The Astrophysical Journal Letters} {\bfseries 907} (2021) L1}.

\bibitem{Matsuoka_2019}
Y.~{Matsuoka}, K.~{Iwasawa}, M.~{Onoue}, N.~{Kashikawa}, M.A.~{Strauss}, C.-H.~{Lee} et~al., \emph{{Subaru High-z Exploration of Low-luminosity Quasars (SHELLQs). X. Discovery of 35 Quasars and Luminous Galaxies at 5.7 {\ensuremath{\leq}} z {\ensuremath{\leq}} 7.0}}, \href{https://doi.org/10.3847/1538-4357/ab3c60}{\emph{\apj} {\bfseries 883} (2019) 183} [\href{https://arxiv.org/abs/1908.07910}{{\ttfamily 1908.07910}}].

\bibitem{witstok24}
J.~{Witstok}, R.~{Maiolino}, R.~{Smit}, G.C.~{Jones}, A.J.~{Bunker}, J.M.~{Helton} et~al., \emph{{JADES: Primeval Lyman-$\mathrm{\alpha}$ emitting galaxies reveal early sites of reionisation out to redshift $z \sim 9$}}, \href{https://doi.org/10.48550/arXiv.2404.05724}{\emph{arXiv e-prints} (2024) arXiv:2404.05724} [\href{https://arxiv.org/abs/2404.05724}{{\ttfamily 2404.05724}}].

\bibitem{mellema2013}
G.~{Mellema}, L.V.E.~{Koopmans}, F.A.~{Abdalla}, G.~{Bernardi}, B.~{Ciardi}, S.~{Daiboo} et~al., \emph{{Reionization and the Cosmic Dawn with the Square Kilometre Array}}, \href{https://doi.org/10.1007/s10686-013-9334-5}{\emph{Experimental Astronomy} {\bfseries 36} (2013) 235} [\href{https://arxiv.org/abs/1210.0197}{{\ttfamily 1210.0197}}].

\bibitem{Majumdar_2011}
S.~{Majumdar}, S.~{Bharadwaj}, K.K.~{Datta} and T.R.~{Choudhury}, \emph{{The impact of anisotropy from finite light traveltime on detecting ionized bubbles in redshifted 21-cm maps}}, \href{https://doi.org/10.1111/j.1365-2966.2011.18223.x}{\emph{\mnras} {\bfseries 413} (2011) 1409} [\href{https://arxiv.org/abs/1006.0430}{{\ttfamily 1006.0430}}].

\bibitem{majumdar2012}
S.~{Majumdar}, S.~{Bharadwaj} and T.R.~{Choudhury}, \emph{{Constrainingquasar and intergalactic medium properties through bubble detection in redshifted 21-cm maps}}, \href{https://doi.org/10.1111/j.1365-2966.2012.21914.x}{\emph{\mnras} {\bfseries 426} (2012) 3178} [\href{https://arxiv.org/abs/1111.6354}{{\ttfamily 1111.6354}}].

\bibitem{ghara2017}
R.~{Ghara}, T.R.~{Choudhury}, K.K.~{Datta} and S.~{Choudhuri}, \emph{{Imaging the redshifted 21 cm pattern around the first sources during the cosmic dawn using the SKA}}, \href{https://doi.org/10.1093/mnras/stw2494}{\emph{\mnras} {\bfseries 464} (2017) 2234} [\href{https://arxiv.org/abs/1607.02779}{{\ttfamily 1607.02779}}].

\bibitem{zack2020}
E.~{Zackrisson}, S.~{Majumdar}, R.~{Mondal}, C.~{Binggeli}, M.~{Sahl{\'e}n}, T.R.~{Choudhury} et~al., \emph{{Bubble mapping with the Square Kilometre Array - I. Detecting galaxies with Euclid, JWST, WFIRST, and ELT within ionized bubbles in the intergalactic medium at z > 6}}, \href{https://doi.org/10.1093/mnras/staa098}{\emph{\mnras} {\bfseries 493} (2020) 855} [\href{https://arxiv.org/abs/1905.00437}{{\ttfamily 1905.00437}}].

\bibitem{giri2018}
S.K.~{Giri}, G.~{Mellema} and R.~{Ghara}, \emph{{Optimal identification of H II regions during reionization in 21-cm observations}}, \href{https://doi.org/10.1093/mnras/sty1786}{\emph{\mnras} {\bfseries 479} (2018) 5596} [\href{https://arxiv.org/abs/1801.06550}{{\ttfamily 1801.06550}}].

\bibitem{giri2018a}
S.K.~{Giri}, G.~{Mellema}, K.L.~{Dixon} and I.T.~{Iliev}, \emph{{Bubble size statistics during reionization from 21-cm tomography}}, \href{https://doi.org/10.1093/mnras/stx2539}{\emph{\mnras} {\bfseries 473} (2018) 2949} [\href{https://arxiv.org/abs/1706.00665}{{\ttfamily 1706.00665}}].

\bibitem{bianco2024}
M.~{Bianco}, S.K.~{Giri}, D.~{Prelogovi{\'c}}, T.~{Chen}, F.G.~{Mertens}, E.~{Tolley} et~al., \emph{{Deep learning approach for identification of H II regions during reionization in 21-cm observations - II. Foreground contamination}}, \href{https://doi.org/10.1093/mnras/stae257}{\emph{\mnras} {\bfseries 528} (2024) 5212} [\href{https://arxiv.org/abs/2304.02661}{{\ttfamily 2304.02661}}].

\bibitem{bianco2024a}
M.~{Bianco}, S.K.~{Giri}, R.~{Sharma}, T.~{Chen}, S.~{Parth Krishna}, C.~{Finlay} et~al., \emph{{Deep learning approach for identification of HII regions during reionization in 21-cm observations -- III. image recovery}}, \href{https://doi.org/10.48550/arXiv.2408.16814}{\emph{arXiv e-prints} (2024) arXiv:2408.16814} [\href{https://arxiv.org/abs/2408.16814}{{\ttfamily 2408.16814}}].

\bibitem{Di_Matteo+2002}
T.~{Di Matteo}, R.~{Perna}, T.~{Abel} and M.J.~{Rees}, \emph{{Radio Foregrounds for the 21 Centimeter Tomography of the Neutral Intergalactic Medium at High Redshifts}}, \href{https://doi.org/10.1086/324293}{\emph{\apj} {\bfseries 564} (2002) 576} [\href{https://arxiv.org/abs/astro-ph/0109241}{{\ttfamily astro-ph/0109241}}].

\bibitem{Oh_2003}
S.P.~{Oh} and K.J.~{Mack}, \emph{{Foregrounds for 21-cm observations of neutral gas at high redshift}}, \href{https://doi.org/10.1111/j.1365-2966.2003.07133.x}{\emph{\mnras} {\bfseries 346} (2003) 871} [\href{https://arxiv.org/abs/astro-ph/0302099}{{\ttfamily astro-ph/0302099}}].

\bibitem{Santos+2005}
M.G.~{Santos}, A.~{Cooray} and L.~{Knox}, \emph{{Multifrequency Analysis of 21 Centimeter Fluctuations from the Era of Reionization}}, \href{https://doi.org/10.1086/429857}{\emph{\apj} {\bfseries 625} (2005) 575} [\href{https://arxiv.org/abs/astro-ph/0408515}{{\ttfamily astro-ph/0408515}}].

\bibitem{Ali_2008}
S.S.~Ali, S.~Bharadwaj and J.N.~Chengalur, \emph{Foregrounds for redshifted 21-cm studies of reionization: Giant meter wave radio telescope 153-mhz observations}, \href{https://doi.org/10.1111/j.1365-2966.2008.12984.x}{\emph{Monthly Notices of the Royal Astronomical Society} {\bfseries 385} (2008) 2166–2174}.

\bibitem{geil2008}
P.M.~{Geil}, J.S.B.~{Wyithe}, N.~{Petrovic} and S.P.~{Oh}, \emph{{The effect of Galactic foreground subtraction on redshifted 21-cm observations of quasar HII regions}}, \href{https://doi.org/10.1111/j.1365-2966.2008.13798.x}{\emph{\mnras} {\bfseries 390} (2008) 1496} [\href{https://arxiv.org/abs/0805.0038}{{\ttfamily 0805.0038}}].

\bibitem{Kakiichi_2017}
K.~{Kakiichi}, S.~{Majumdar}, G.~{Mellema}, B.~{Ciardi}, K.L.~{Dixon}, I.T.~{Iliev} et~al., \emph{{Recovering the H II region size statistics from 21-cm tomography}}, \href{https://doi.org/10.1093/mnras/stx1568}{\emph{\mnras} {\bfseries 471} (2017) 1936} [\href{https://arxiv.org/abs/1702.02520}{{\ttfamily 1702.02520}}].

\bibitem{Datta_2007}
K.K.~Datta, S.~Bharadwaj and T.R.~Choudhury, \emph{Detecting ionized bubbles in redshifted 21-cm maps}, \href{https://doi.org/10.1111/j.1365-2966.2007.12421.x}{\emph{Monthly Notices of the Royal Astronomical Society} {\bfseries 382} (2007) 809}.

\bibitem{Datta_2008}
K.K.~Datta, S.~Majumdar, S.~Bharadwaj and T.R.~Choudhury, \emph{Simulating the impact of h{\hspace{1em} }i fluctuations on matched filter search for ionized bubbles in redshifted 21-cm maps}, \href{https://doi.org/10.1111/j.1365-2966.2008.14008.x}{\emph{Monthly Notices of the Royal Astronomical Society} {\bfseries 391} (2008) 1900}.

\bibitem{Datta_2012}
K.K.~{Datta}, M.M.~{Friedrich}, G.~{Mellema}, I.T.~{Iliev} and P.R.~{Shapiro}, \emph{{Prospects of observing a quasar H II region during the epoch of reionization with the redshifted 21-cm signal}}, \href{https://doi.org/10.1111/j.1365-2966.2012.21268.x}{\emph{\mnras} {\bfseries 424} (2012) 762} [\href{https://arxiv.org/abs/1203.0517}{{\ttfamily 1203.0517}}].

\bibitem{Datta_2009}
K.K.~Datta, S.~Bharadwaj and T.R.~Choudhury, \emph{The optimal redshift for detecting ionized bubbles in h<scp>i</scp> 21-cm maps}, \href{https://doi.org/10.1111/j.1745-3933.2009.00739.x}{\emph{Monthly Notices of the Royal Astronomical Society: Letters} {\bfseries 399} (2009) L132–L136}.

\bibitem{Ghara_2020}
R.~{Ghara} and T.R.~{Choudhury}, \emph{{Bayesian approach to constraining the properties of ionized bubbles during reionization}}, \href{https://doi.org/10.1093/mnras/staa1599}{\emph{\mnras} {\bfseries 496} (2020) 739} [\href{https://arxiv.org/abs/1909.12317}{{\ttfamily 1909.12317}}].

\bibitem{mesinger2016}
A.~{Mesinger}, B.~{Greig} and E.~{Sobacchi}, \emph{{The Evolution Of 21 cm Structure (EOS): public, large-scale simulations of Cosmic Dawn and reionization}}, \href{https://doi.org/10.1093/mnras/stw831}{\emph{\mnras} {\bfseries 459} (2016) 2342} [\href{https://arxiv.org/abs/1602.07711}{{\ttfamily 1602.07711}}].

\bibitem{majumdar2018}
S.~{Majumdar}, J.R.~{Pritchard}, R.~{Mondal}, C.A.~{Watkinson}, S.~{Bharadwaj} and G.~{Mellema}, \emph{{Quantifying the non-Gaussianity in the EoR 21-cm signal through bispectrum}}, \href{https://doi.org/10.1093/mnras/sty535}{\emph{\mnras} {\bfseries 476} (2018) 4007} [\href{https://arxiv.org/abs/1708.08458}{{\ttfamily 1708.08458}}].

\bibitem{majumdar2020}
S.~{Majumdar}, M.~{Kamran}, J.R.~{Pritchard}, R.~{Mondal}, A.~{Mazumdar}, S.~{Bharadwaj} et~al., \emph{{Redshifted 21-cm bispectrum - I. Impact of the redshift space distortions on the signal from the Epoch of Reionization}}, \href{https://doi.org/10.1093/mnras/staa3168}{\emph{\mnras} {\bfseries 499} (2020) 5090} [\href{https://arxiv.org/abs/2007.06584}{{\ttfamily 2007.06584}}].

\bibitem{trott2019}
C.M.~{Trott}, C.A.~{Watkinson}, C.H.~{Jordan}, S.~{Yoshiura}, S.~{Majumdar}, N.~{Barry} et~al., \emph{{Gridded and direct Epoch of Reionisation bispectrum estimates using the Murchison Widefield Array}}, \href{https://doi.org/10.1017/pasa.2019.15}{\emph{\pasa} {\bfseries 36} (2019) e023} [\href{https://arxiv.org/abs/1905.07161}{{\ttfamily 1905.07161}}].

\bibitem{ghara2020}
R.~{Ghara}, S.K.~{Giri}, G.~{Mellema}, B.~{Ciardi}, S.~{Zaroubi}, I.T.~{Iliev} et~al., \emph{{Constraining the intergalactic medium at z {\ensuremath{\approx}} 9.1 using LOFAR Epoch of Reionization observations}}, \href{https://doi.org/10.1093/mnras/staa487}{\emph{\mnras} {\bfseries 493} (2020) 4728} [\href{https://arxiv.org/abs/2002.07195}{{\ttfamily 2002.07195}}].

\bibitem{kamran2021}
M.~{Kamran}, R.~{Ghara}, S.~{Majumdar}, R.~{Mondal}, G.~{Mellema}, S.~{Bharadwaj} et~al., \emph{{Redshifted 21-cm bispectrum - II. Impact of the spin temperature fluctuations and redshift space distortions on the signal from the Cosmic Dawn}}, \href{https://doi.org/10.1093/mnras/stab216}{\emph{\mnras} {\bfseries 502} (2021) 3800} [\href{https://arxiv.org/abs/2012.11616}{{\ttfamily 2012.11616}}].

\bibitem{noble2024}
L.~{Noble}, M.~{Kamran}, S.~{Majumdar}, C.~{Shekhar Murmu}, R.~{Ghara}, G.~{Mellema} et~al., \emph{{Impact of the Epoch of Reionization sources on the 21-cm bispectrum}}, \href{https://doi.org/10.48550/arXiv.2406.03118}{\emph{arXiv e-prints} (2024) arXiv:2406.03118} [\href{https://arxiv.org/abs/2406.03118}{{\ttfamily 2406.03118}}].

\bibitem{Bennett_2013}
C.L.~Bennett, D.~Larson, J.L.~Weiland, N.~Jarosik, G.~Hinshaw, N.~Odegard et~al., \emph{Nine-year wilkinson microwave anisotropy probe ( wmap ) observations: Final maps and results}, \href{https://doi.org/10.1088/0067-0049/208/2/20}{\emph{The Astrophysical Journal Supplement Series} {\bfseries 208} (2013) 20}.

\bibitem{Bharadwaj_2004}
S.~Bharadwaj and S.S.~Ali, \emph{The cosmic microwave background radiation fluctuations from h{\hspace{1em} }i perturbations prior to reionization}, \href{https://doi.org/10.1111/j.1365-2966.2004.07907.x}{\emph{Monthly Notices of the Royal Astronomical Society} {\bfseries 352} (2004) 142}.

\bibitem{Bharadwaj_2005}
S.~Bharadwaj and S.S.~Ali, \emph{On using visibility correlations to probe the h{\hspace{1em} }i distribution from the dark ages to the present epoch - i. formalism and the expected signal}, \href{https://doi.org/10.1111/j.1365-2966.2004.08604.x}{\emph{Monthly Notices of the Royal Astronomical Society} {\bfseries 356} (2005) 1519}.

\bibitem{Wouthuysen+1952}
S.A.~{Wouthuysen}, \emph{{On the excitation mechanism of the 21-cm (radio-frequency) interstellar hydrogen emission line.}}, \href{https://doi.org/10.1086/106661}{\emph{The Astronomical Journal} {\bfseries 57} (1952) 31}.

\bibitem{Field+1958}
G.B.~{Field}, \emph{{Excitation of the Hydrogen 21-CM Line}}, \href{https://doi.org/10.1109/JRPROC.1958.286741}{\emph{Proceedings of the IRE} {\bfseries 46} (1958) 240}.

\bibitem{Madau_1997}
P.~{Madau}, A.~{Meiksin} and M.J.~{Rees}, \emph{{21 Centimeter Tomography of the Intergalactic Medium at High Redshift}}, \href{https://doi.org/10.1086/303549}{\emph{\apj} {\bfseries 475} (1997) 429} [\href{https://arxiv.org/abs/astro-ph/9608010}{{\ttfamily astro-ph/9608010}}].

\bibitem{Bharadwaj_2004a}
S.~Bharadwaj and P.S.~Srikant, \emph{{HI} fluctuations at large redshifts: {III} {\textemdash} simulating the signal expected at {GMRT}}, \href{https://doi.org/10.1007/bf02702289}{\emph{Journal of Astrophysics and Astronomy} {\bfseries 25} (2004) 67}.

\bibitem{Mondal_2015}
R.~Mondal, S.~Bharadwaj, S.~Majumdar, A.~Bera and A.~Acharyya, \emph{{The effect of non-Gaussianity on error predictions for the Epoch of Reionization (EoR) 21-cm power spectrum}}, \href{https://doi.org/10.1093/mnrasl/slv015}{\emph{Monthly Notices of the Royal Astronomical Society: Letters} {\bfseries 449} (2015) L41} [\href{https://arxiv.org/abs/https://academic.oup.com/mnrasl/article-pdf/449/1/L41/9419984/slv015.pdf}{{\ttfamily https://academic.oup.com/mnrasl/article-pdf/449/1/L41/9419984/slv015.pdf}}].

\bibitem{Choudhury_2009}
T.R.~Choudhury, M.G.~Haehnelt and J.~Regan, \emph{{Inside-out or outside-in: the topology of reionization in the photon-starved regime suggested by Ly-alpha forest data}}, \href{https://doi.org/10.1111/j.1365-2966.2008.14383.x}{\emph{Monthly Notices of the Royal Astronomical Society} {\bfseries 394} (2009) 960} [\href{https://arxiv.org/abs/https://academic.oup.com/mnras/article-pdf/394/2/960/3710519/mnras0394-0960.pdf}{{\ttfamily https://academic.oup.com/mnras/article-pdf/394/2/960/3710519/mnras0394-0960.pdf}}].

\bibitem{Majumdar_2014}
S.~Majumdar, G.~Mellema, K.K.~Datta, H.~Jensen, T.R.~Choudhury, S.~Bharadwaj et~al., \emph{{On the use of seminumerical simulations in predicting the 21-cm signal from the epoch of reionization}}, \href{https://doi.org/10.1093/mnras/stu1342}{\emph{Monthly Notices of the Royal Astronomical Society} {\bfseries 443} (2014) 2843} [\href{https://arxiv.org/abs/https://academic.oup.com/mnras/article-pdf/443/4/2843/6274877/stu1342.pdf}{{\ttfamily https://academic.oup.com/mnras/article-pdf/443/4/2843/6274877/stu1342.pdf}}].

\bibitem{Mondal_2017}
R.~{Mondal}, S.~{Bharadwaj} and S.~{Majumdar}, \emph{{Statistics of the epoch of reionization (EoR) 21-cm signal - II. The evolution of the power-spectrum error-covariance}}, \href{https://doi.org/10.1093/mnras/stw2599}{\emph{\mnras} {\bfseries 464} (2017) 2992} [\href{https://arxiv.org/abs/1606.03874}{{\ttfamily 1606.03874}}].

\bibitem{Jeli__2008}
V.~Jeli{\'{c} }, S.~Zaroubi, P.~Labropoulos, R.M.~Thomas, G.~Bernardi, M.A.~Brentjens et~al., \emph{Foreground simulations for the {LOFAR}-epoch of reionization experiment}, \href{https://doi.org/10.1111/j.1365-2966.2008.13634.x}{\emph{Monthly Notices of the Royal Astronomical Society} {\bfseries 389} (2008) 1319}.

\bibitem{Choudhuri_2014}
S.~Choudhuri, S.~Bharadwaj, A.~Ghosh and S.S.~Ali, \emph{Visibility-based angular power spectrum estimation in low-frequency radio interferometric observations}, \href{https://doi.org/10.1093/mnras/stu2027}{\emph{Monthly Notices of the Royal Astronomical Society} {\bfseries 445} (2014) 4351}.

\bibitem{10.1111/j.1365-2966.2012.21889.x}
A.~Ghosh, J.~Prasad, S.~Bharadwaj, S.S.~Ali and J.N.~Chengalur, \emph{{Characterizing foreground for redshifted 21 cm radiation: 150 MHz Giant Metrewave Radio Telescope observations}}, \href{https://doi.org/10.1111/j.1365-2966.2012.21889.x}{\emph{Monthly Notices of the Royal Astronomical Society} {\bfseries 426} (2012) 3295} [\href{https://arxiv.org/abs/https://academic.oup.com/mnras/article-pdf/426/4/3295/3332953/426-4-3295.pdf}{{\ttfamily https://academic.oup.com/mnras/article-pdf/426/4/3295/3332953/426-4-3295.pdf}}].

\bibitem{2017CSci..113..707G}
Y.~{Gupta}, B.~{Ajithkumar}, H.S.~{Kale}, S.~{Nayak}, S.~{Sabhapathy}, S.~{Sureshkumar} et~al., \emph{{The upgraded GMRT: opening new windows on the radio Universe}}, \href{https://doi.org/10.18520/cs/v113/i04/707-714}{\emph{Current Science} {\bfseries 113} (2017) 707}.

\bibitem{ska_telescope}
{SKA Observatory}, ``Ska telescope specifications.'' \url{https://www.skao.int/en/science-users/118/ska-telescope-specifications#:~:text=SKA1%2DLow%20(also%20referred%20to,stations%20of%20256%20antennas%20each.}, 2024.

\bibitem{ali2005}
S.~{Bharadwaj} and S.S.~{Ali}, \emph{{On using visibility correlations to probe the HI distribution from the dark ages to the present epoch - I. Formalism and the expected signal}}, \href{https://doi.org/10.1111/j.1365-2966.2004.08604.x}{\emph{\mnras} {\bfseries 356} (2005) 1519} [\href{https://arxiv.org/abs/astro-ph/0406676}{{\ttfamily astro-ph/0406676}}].

\bibitem{Cho_2010}
J.~Cho and A.~Lazarian, \emph{Galactic foregrounds: Spatial fluctuations and a procedure for removal}, \href{https://doi.org/10.1088/0004-637X/720/2/1181}{\emph{The Astrophysical Journal} {\bfseries 720} (2010) 1181}.

\bibitem{Martire_2022}
F.~Martire, R.~Barreiro and E.~Martínez-González, \emph{Characterization of the polarized synchrotron emission from planck and wmap data}, \href{https://doi.org/10.1088/1475-7516/2022/04/003}{\emph{Journal of Cosmology and Astroparticle Physics} {\bfseries 2022} (2022) 003}.

\bibitem{saikat24}
S.~{Gayen}, R.~{Sagar}, S.~{Mangla}, P.~{Dutta}, N.~{Roy}, A.~{Chakraborty} et~al., \emph{{Calibration requirement for Epoch of Reionization 21-cm signal observation. Part III. Bias and variance in uGMRT ELAIS-N1 field power spectrum}}, \href{https://doi.org/10.1088/1475-7516/2024/05/068}{\emph{\jcap} {\bfseries 2024} (2024) 068} [\href{https://arxiv.org/abs/2402.18306}{{\ttfamily 2402.18306}}].

\end{thebibliography}\endgroup

\end{document}